\documentclass[a4paper,12pt,onecolumn,nofootinbib,preprint]{article}
\usepackage{amssymb,amsmath,epsfig,subfig} 
\usepackage{times}
\usepackage{cite}
\usepackage{subfloat}
\usepackage{hyperref}
\usepackage{graphicx}
\usepackage{url}
\usepackage{color}
\def\baselinestretch{1.4}
\begin{document}
 \topmargin -0.1in
\headsep 30pt
\footskip 40pt
\oddsidemargin 12pt
\evensidemargin -16pt
\textheight 8.5in
\textwidth 6.5in
\parindent 20pt
 
\def\baselinestretch{1.5}
\newcommand{\newc}{\newcommand}
\def\preprint{{preprint}}
\def\Ord{\lower .7ex\hbox{$\;\stackrel{\textstyle <}{\sim}\;$}}
\def\OOrd{\lower .7ex\hbox{$\;\stackrel{\textstyle >}{\sim}\;$}}
\def\cO#1{{\cal{O}}\left(#1\right)}
\newc{\order}{{\cal O}}
\def\lag             {{\cal L}}
\def\Lag             {{\cal L}}
\def\lum             {{\cal L}}
\def\R               {{\cal R}}
\def\Rsq             {{\cal R}^{\sq}}
\def\Rst             {{\cal R}^{\st}}
\def\Rsb             {{\cal R}^{\sb}}
\def\M               {{\cal M}}
\def\Oas             {{\cal O}(\alpha_{s})}
\def\Vcal            {{\cal V}}
\def\Wcal            {{\cal W}}
\newc{\be}{\begin{equation}}
\newc{\ee}{\end{equation}}
\newc{\br}{\begin{eqnarray}}
\newc{\er}{\end{eqnarray}}
\newc{\ba}{\begin{array}}
\newc{\ea}{\end{array}}
\newc{\bi}{\begin{itemize}}
\newc{\ei}{\end{itemize}}
\newc{\bn}{\begin{enumerate}}
\newc{\en}{\end{enumerate}}
\newc{\bc}{\begin{center}}
\newc{\ec}{\end{center}}
\newc{\ul}{\underline}
\newc{\ra}{\rightarrow}
\newc{\lra}{\longrightarrow}
\newc{\wt}{\widetilde}
\newc{\til}{\tilde}
\def\kr              {^{\dagger}}
\newc{\wh}{\widehat}
\newc{\ti}{\times}
\newc{\Dir}{\kern -6.4pt\Big{/}}
\newc{\Dirin}{\kern -10.4pt\Big{/}\kern 4.4pt}
\newc{\DDir}{\kern -10.6pt\Big{/}}
\newc{\DGir}{\kern -6.0pt\Big{/}}
\newc{\sig}{\sigma}
\newc{\sigmalstop}{\sig_{\lstoppair}}
\newc{\Sig}{\Sigma}  
\newc{\del}{\delta}
\newc{\Del}{\Delta}
\newc{\lam}{\lambda}
\newc{\Lam}{\Lambda}
\newc{\gam}{\gamma}
\newc{\Gam}{\Gamma}
\newc{\eps}{\epsilon}
\newc{\Eps}{\Epsilon}
\newc{\kap}{\kappa}
\newc{\Kap}{\Kappa}
\newc{\modulus}[1]{\left| #1 \right|}
\newc{\eq}[1]{(\ref{eq:#1})}
\newc{\eqs}[2]{(\ref{eq:#1},\ref{eq:#2})}
\newc{\etal}{{\it et al.}\ }
\newc{\ibid}{{\it ibid}.}
\newc{\ibidem}{{\it ibidem}.}
\newc{\eg}{{\it e.g.}\ }
\newc{\ie}{{\it i.e.}\ }
\def \viz{\emph{viz.}}
\def \etc{\emph{etc. }}
\newc{\nonum}{\nonumber}
\newc{\lab}[1]{\label{eq:#1}}
\newc{\dpr}[2]{({#1}\cdot{#2})}
\newc{\lt}{\stackrel{<}}
\newc{\gt}{\stackrel{>}}
\newc{\lsimeq}{\stackrel{<}{\sim}}
\newc{\gsimeq}{\stackrel{>}{\sim}}
\def\lsim{\buildrel{\scriptscriptstyle <}\over{\scriptscriptstyle\sim}}
\def\gsim{\buildrel{\scriptscriptstyle >}\over{\scriptscriptstyle\sim}}
\def\lapp{\mathrel{\rlap{\raise.5ex\hbox{$<$}}
                    {\lower.5ex\hbox{$\sim$}}}}
\def\gapp{\mathrel{\rlap{\raise.5ex\hbox{$>$}}
                    {\lower.5ex\hbox{$\sim$}}}}
\newc{\half}{\frac{1}{2}}
\newcommand {\nnc}        {{\overline{\mathrm N}_{95}}}
\newcommand {\dm}         {\Delta m}
\newcommand {\dM}         {\Delta M}
\def\bra{\langle}
\def\ket{\rangle}
\def\cO#1{{\cal{O}}\left(#1\right)}
\def \DM{{\Delta{m}}}
\newc{\bQ}{\ol{Q}}
\newc{\dota}{\dot{\alpha }}
\newc{\dotb}{\dot{\beta }}
\newc{\dotd}{\dot{\delta }}
\newc{\nindnt}{\noindent}

\newcommand{\medf}[2] {{\footnotesize{\frac{#1}{#2}} }}
\newcommand{\smaf}[2] {{\textstyle \frac{#1}{#2} }}
\def\onesq            {{\textstyle \frac{1}{\sqrt{2}} }}
\def\onehf            {{\textstyle \frac{1}{2} }}
\def\oneth            {{\textstyle \frac{1}{3} }}
\def\twoth            {{\textstyle \frac{2}{3} }}
\def\onefo            {{\textstyle \frac{1}{4} }}
\def\forth            {{\textstyle \frac{4}{3} }}

\newc{\matth}{\mathsurround=0pt}
\def\ML{\ifmmode{{\mathaccent"7E M}_L}
             \else{${\mathaccent"7E M}_L$}\fi}
\def\MR{\ifmmode{{\mathaccent"7E M}_R}
             \else{${\mathaccent"7E M}_R$}\fi}
\newcommand{\s}{\\ \vspace*{-3mm} }

\def \ud { {1 \over 2} }
\def \ut { {1 \over 3} }
\def \td { {3 \over 2} }
\newc{\mr}{\mathrm}
\def\dh {\partial }
\def \cs { cross-section }
\def \css { cross-sections }
\def \cm { centre of mass }
\def \cms { centre of mass energy }
\def \cc { coupling constant }
\def \ccs {coupling constants }
\def \gc {gauge coupling }
\def \gcc {gauge coupling constant }
\def \gccs {gauge coupling constants }
\def \yc {Yukawa coupling }
\def \ycc {Yukawa coupling constant }
\def \pp {{parameter }}
\def \pps {{parameters }} 
\def \ps {parameter space }
\def \pss {parameter spaces }
\def \vv {vice versa }

\newc{\siminf}{\mbox{$_{\sim}$ {\small {\hspace{-1.em}{$<$}}}    }}
\newc{\simsup}{\mbox{$_{\sim}$ {\small {\hspace{-1.em}{$>$}}}    }}


\newc {\Zboson}{{\mathrm Z}^{0}}
\newc{\thetaw}{\theta_W}
\newc{\mbot}{{m_b}}
\newc{\mtop}{{m_t}}
\newc{\sm}{${\cal {SM}}$}
\newc{\as}{\alpha_s}
\newc{\aem}{\alpha_{em}}
\def \PI{{\pi^{\pm}}}
\newc{\ppbar}{\mbox{$p\ol{p}$}}
\newc{\bbbar}{\mbox{$b\ol{b}$}}
\newc{\ccbar}{\mbox{$c\ol{c}$}}
\newc{\ttbar}{\mbox{$t\ol{t}$}}
\newc{\eebar}{\mbox{$e\ol{e}$}}
\newc{\zzero}{\mbox{$Z^0$}}
\def \gamz{\Gam_Z}
\newc{\wplus}{\mbox{$W^+$}}
\newc{\wminus}{\mbox{$W^-$}}
\newc{\ellp}{\ell^+}
\newc{\ellm}{\ell^-}
\newc{\elp}{\mbox{$e^+$}}
\newc{\elm}{\mbox{$e^-$}}
\newc{\elpm}{\mbox{$e^{\pm}$}}
\newc{\qbar}     {\mbox{$\ol{q}$}}
\def \ewgroup{SU(2)_L \otimes U(1)_Y}
\def \smgroup{SU(3)_C \otimes SU(2)_L \otimes U(1)_Y}
\def \smcolorem{SU(3)_C \otimes U(1)_{em}}

\def \SSM  {Supersymmetric Standard Model}
\def \poincare{Poincare$\acute{e}$}
\def \superspace{\emph{superspace}}
\def \sfs{\emph{superfields}}
\def \superpot{\emph{superpotential}}
\def \csf{\emph{chiral superfield}}
\def \csfs{\emph{chiral superfields}}
\def \vsf{\emph{vector superfield }}
\def \vsfs{\emph{vector superfields}}
\newc{\Ebar}{{\bar E}}
\newc{\Dbar}{{\bar D}}
\newc{\Ubar}{{\bar U}}
\newc{\susy}{{{SUSY}}}
\newc{\msusy}{{{M_{SUSY}}}}

\def\photino{\ifmmode{\mathaccent"7E \gam}\else{$\mathaccent"7E \gam$}\fi}
\def\taugluino{\ifmmode{\tau_{\mathaccent"7E g}}
             \else{$\tau_{\mathaccent"7E g}$}\fi}
\def\mphotino{\ifmmode{m_{\mathaccent"7E \gam}}
             \else{$m_{\mathaccent"7E \gam}$}\fi}
\newc{\gl}   {\mbox{$\wt{g}$}}
\newc{\mgl}  {\mbox{$m_{\gl}$}}
\def \charginopm{{\wt\chi}^{\pm}}
\def \mcharginopm{m_{\charginopm}}
\def \mchpmmin {\mcharginopm^{min}}
\def \chonep {{\wt\chi_1^+}}
\def \chone {{\wt\chi_1}}
\def \ch2p {{\wt\chi_2^+}}
\def \chonem {{\wt\chi_1^-}}
\def \ch2m {{\wt\chi_2^-}}
\def \chplus {{\wt\chi^+}}
\def \chminus {{\wt\chi^-}}
\def \chonip{{\wt\chi_i}^{+}}
\def \chonim{{\wt\chi_i}^{-}}
\def \chonipm{{\wt\chi_i}^{\pm}}
\def \chonjp{{\wt\chi_j}^{+}}
\def \chonjm{{\wt\chi_j}^{-}}
\def \chonjpm{{\wt\chi_j}^{\pm}}
\def \chonepm{{\wt\chi_1}^{\pm}}
\def \chonemp{{\wt\chi_1}^{\mp}}
\def \mchonepm{m_{\chonepm}}
\def \mchonemp{m_{\chonemp}}
\def \chtwopm{{\wt\chi_2}^{\pm}}
\def \mchtwopm{m_{\chtwopm}}
\newc{\dmchi}{\Delta m_{\wt\chi}}


\def \vlsp{\emph{VLSP}}
\def \lspi{\wt\chi_i^0}
\def \mlspi{m_{\lspi}}
\def \lspj{\wt\chi_j^0}
\def \mlspj{m_{\lspj}}
\def \lspone{\wt\chi_1^0}
\def \mlspone{m_{\lspone}}
\def \lsptwo{\wt\chi_2^0}
\def \mlsptwo{m_{\lsptwo}}
\def \lspthree{\wt\chi_3^0}
\def \mlspthree{m_{\lspthree}}
\def \lspfour{\wt\chi_4^0}
\def \mlspfour{m_{\lspfour}}


\newc{\sele}{\wt{\mathrm e}}
\newc{\sell}{\wt{\ell}}
\def \msell{m_{\sell}}
\def \slepone{\wt\ell_1}
\def \mslepone{m_{\slepone}}
\def \smuone{\wt\mu_1}
\def \msmuone{m_{\smuone}}
\def \stauone{\wt\tau_1}
\def \stauonepm{{\wt\tau_1}^\pm}
\def \mstauone{m_{\stauone}}
\def \snu{\wt{\nu}}
\def \snutau{\wt{\nu}_{\tau}}
\def \nutau{{\nu}_{\tau}}                  
\def \msnu{m_{\snu}}
\def \msnumu{m_{\snu_{\mu}}}
\def \barsnu{\wt{\bar{\nu}}}
\def \barsnul{\barsnu_{\ell}}
\def \snul{\snu_{\ell}}
\def \mbarsnu{m_{\barsnu}}
\newc{\snue}     {\mbox{$ \wt{\nu_e}$}}
\newc{\smu}{\wt{\mu}}
\newc{\stau}{\wt{\tau}}
\newc {\nuL} {\wt{\nu}_L}
\newc {\nuR} {\wt{\nu}_R}
\newc {\snub} {\bar{\wt{\nu}}}
\newc {\eL} {\wt{e}_L}
\newc {\eR} {\wt{e}_R}
\def \slepl{\wt{l}_L}
\def \mslepl{m_{\slepl}}
\def \slepr{\wt{l}_R}
\def \mslepr{m_{\slepr}}
\def \stau{\wt\tau}
\def \mstau{m_{\stau}}
\def \slepton{\wt\ell}
\def \mslepton{m_{\slepton}}
\def \mlhiggs{m_{h^0}}

\def \xr{X_{r}}

\def \sfer{\wt{f}}
\def \msfer{m_{\sfer}}
\def \sq{\wt{q}}
\def \msq{m_{\sq}}
\def \msquleft{m_{\tilde{u_L}}}
\def \msqurht{m_{\tilde{u_R}}}
\def \sql{\wt{q}_L}
\def \msql{m_{\sql}}
\def \sqr{\wt{q}_R}
\def \msqr{m_{\sqr}}
\newc{\msqot}  {\mbox{$m_(\sq_{1,2} )$}}
\newc{\sqbar}    {\mbox{$\bar{\wt{q}}$}}
\newc{\ssb}      {\mbox{$\squark\ol{\squark}$}}
\newc {\qL} {\wt{q}_L}
\newc {\qR} {\wt{q}_R}
\newc {\uL} {\wt{u}_L}
\newc {\uR} {\wt{u}_R}
\def \ul{\wt{u}_L}
\def \mul{m_{\ul}}
\newc {\dL} {\wt{d}_L}
\newc {\dR} {\wt{d}_R}
\newc {\cL} {\wt{c}_L}
\newc {\cR} {\wt{c}_R}
\newc {\sL} {\wt{s}_L}
\newc {\sR} {\wt{s}_R}
\newc {\tL} {\wt{t}_L}
\newc {\tR} {\wt{t}_R}
\newc {\stb} {\ol{\wt{t}}_1}
\newc {\sbot} {\wt{b}_1}
\newc {\msbot} {m_{\sbot}}
\newc {\sbotb} {\ol{\wt{b}}_1}
\newc {\bL} {\wt{b}_L}
\newc {\bR} {\wt{b}_R}
\def \mul{m_{\wt{u}_L}}
\def \mur{m_{\wt{u}_R}}
\def \mdl{m_{\wt{d}_L}}
\def \mdr{m_{\wt{d}_R}}
\def \mcl{m_{\wt{c}_L}}
\def \charml{\wt{c}_L}
\def \mcr{m_{\wt{c}_R}}
\newc{\csquark}  {\mbox{$\wt{c}$}}
\newc{\csquarkl} {\mbox{$\wt{c}_L$}}
\newc{\mcsl}     {\mbox{$m(\csquarkl)$}}
\def \msl{m_{\wt{s}_L}}
\def \msr{m_{\wt{s}_R}}
\def \mbl{m_{\wt{b}_L}}
\def \mbr{m_{\wt{b}_R}}
\def \mtl{m_{\wt{t}_L}}
\def \mtr{m_{\wt{t}_R}}
\def \st{\wt{t}}
\def \mst{m_{\st}}
\newc {\stopl}         {\wt{\mathrm{t}}_{\mathrm L}}
\newc {\stopr}         {\wt{\mathrm{t}}_{\mathrm R}}
\newc {\stoppair}      {\wt{\mathrm{t}}_{1}
\bar{\wt{\mathrm{t}}}_{1}}
\def \lstop{\wt{t}_{1}}
\def \lstopbar{\lstop^*}
\def \hstop{\wt{t}_{2}}
\def \hstopbar{\hstop^*}
\def \mlstop{m_{\lstop}}
\def \mhstop{m_{\hstop}}
\def \lstoppair{\lstop\lstop^*}
\def \hstoppair{\hstop\hstop^*}
\newc{\tsquark}  {\mbox{$\wt{t}$}}
\newc{\ttb}      {\mbox{$\tsquark\ol{\tsquark}$}}
\newc{\ttbone}   {\mbox{$\tsquark_1\ol{\tsquark}_1$}}
\def \tsq {top squark }
\def \tsqs {top squarks }
\def \tsql {ligtest top squark }
\def \tsqh {heaviest top squark }
\newc{\mix}{\theta_{\wt t}}
\newc{\cost}{\cos{\theta_{\wt t}}}
\newc{\sint}{\sin{\theta_{\wt t}}}
\newc{\costloop}{\cos{\theta_{\wt t_{loop}}}}
\def \lsbot{\wt{b}_{1}}
\def \lsbotbar{\lsbot^*}
\def \hsbot{\wt{b}_{2}}
\def \hsbotbar{\hsbot^*}
\def \mlsbot{m_{\lsbot}}
\def \mhsbot{m_{\hsbot}}
\def \lsbotpair{\lsbot\lsbot^*}
\def \hsbotpair{\hsbot\hsbot^*}
\newc{\mixsbot}{\theta_{\wt b}}

\def \mhone{m_{h_1}}
\def \hup{{H_u}}
\def \hdn{{H_d}}
\newc{\tb}{\tan\beta}
\newc{\cb}{\cot\beta}
\newc{\vev}[1]{{\left\langle #1\right\rangle}}

\def \abot{A_{b}}
\def \atop{A_{t}}
\def \atau{A_{\tau}}
\newc{\mhalf}{m_{1/2}}
\newc{\mzero} {\mbox{$m_0$}}
\newc{\azero} {\mbox{$A_0$}}

\newc{\lb}{\lam}
\newc{\lbp}{\lam^{\prime}}
\newc{\lbpp}{\lam^{\prime\prime}}
\newc{\rpv}{{\not \!\! R_p}}
\newc{\rpvm}{{\not  R_p}}
\newc{\rp}{R_{p}}
\newc{\rpmssm}{{RPC MSSM}}
\newc{\rpvmssm}{{RPV MSSM}}


\newc{\sbyb}{S/$\sqrt B$}
\newc{\pelp}{\mbox{$e^+$}}
\newc{\pelm}{\mbox{$e^-$}}
\newc{\pelpm}{\mbox{$e^{\pm}$}}
\newc{\epem}{\mbox{$e^+e^-$}}
\newc{\lplm}{\mbox{$\ell^+\ell^-$}}
\def \branch{\emph{BR}}
\def \branche{\branch(\lstop\ra be^{+}\nu_e \lspone)\ti \branch(\lstop^{*}\ra \bar{b}q\bar{q^{\prime}}\lspone)}
\def \branchmu{\branch(\lstop\ra b\mu^{+}\nu_{\mu} \lspone)\ti \branch(\lstop^{*}\ra \bar{b}q\bar{q^{\prime}}\lspone)}
\def\Ecm{\ifmmode{E_{\mathrm{cm}}}\else{$E_{\mathrm{cm}}$}\fi}
\newc{\rts}{\sqrt{s}}
\newc{\rtshat}{\sqrt{\hat s}}
\newc{\gev}{\,GeV}
\newc{\mev}{~{\rm MeV}}
\newc{\tev}  {\mbox{$\;{\rm TeV}$}}
\newc{\gevc} {\mbox{$\;{\rm GeV}/c$}}
\newc{\gevcc}{\mbox{$\;{\rm GeV}/c^2$}}
\newc{\intlum}{\mbox{${ \int {\cal L} \; dt}$}}
\newc{\call}{{\cal L}}
\def \met  {\mbox{${E\!\!\!\!/_T}$}}
\def \cpv  {\mbox{${CP\!\!\!\!/}$}}
\newc{\ptmiss}{/ \hskip-7pt p_T}
\def \eslash{\not \! E}
\def \etslash{\not \! E_T }
\def \ptslash{\not \! p_T }
\newc{\PT}{\mbox{$p_T$}}
\newc{\ET}{\mbox{$E_T$}}
\newc{\dedx}{\mbox{${\rm d}E/{\rm d}x$}}
\newc{\ifb}{\mbox{${\rm fb}^{-1}$}}
\newc{\ipb}{\mbox{${\rm pb}^{-1}$}}
\newc{\pb}{~{\rm pb}}
\newc{\fb}{~{\rm fb}}
\newc{\ycut}{y_{\mathrm{cut}}}
\newc{\chis}{\mbox{$\chi^{2}$}}
\def \hadron{\emph{hadron}}
\def \nlc{\emph{NLC }}
\def \lhc{\emph{LHC }}
\def \cdf{\emph{CDF }}
\def\dzero{\emptyset}
\def \tevatron{\emph{Tevatron }}
\def \lep{\emph{LEP }}
\def \jets{\emph{jets }}
\def \jet(s){\emph{jet(s) }}

\def\Crs{stroke [] 0 setdash exch hpt sub exch vpt add hpt2 vpt2 neg V currentpoint stroke 
hpt2 neg 0 R hpt2 vpt2 V stroke}
\def\loopdk{\lstop \ra c \lspone}
\def\brloopdk{\branch(\loopdk)}
\def\fourdk{\lstop \ra b \lspone  f \bar f'}
\def\brfourdk{\branch(\fourdk)}
\def\fourdklep{\lstop \ra b \lspone  \ell \nu_{\ell}}
\def\fourdkhad{\lstop \ra b \lspone  q \bar q'}
\def\brfourdklep{\branch(\fourdklep)}
\def\brfourdkhad{\branch(\fourdkhad)}
\def\twodk{\lstop \ra b \chonep}
\def\brtwodk{\branch(\twodk)}
\def\threedkslep{\lstop \ra b \wt{\ell} \nu_{\ell}}
\def\brthreedkslep{\branch(\threedkslep)}
\def\threedksnu{\lstop \ra b \wt{\nu_{\ell}} \ell }
\def\brthreedksnu{\branch(\threedksnu) }
\def\threedklsp{\lstop \ra b W \lspone }
\def\brthreedklsp{\\branch(\threedklsp) }
\def\topdk{t \ra \lstop \lspone}
\def\rpvdk{\lstop \ra e^+ d}
\def\brrpvdk{\branch(\rpvdk)}
\def\fonec{f_{11c}} 
\newc{\mpl}{M_{\rm Pl}}
\newc{\mgut}{M_{GUT}}
\newc{\mw}{M_{W}}
\newc{\mweak}{M_{weak}}
\newc{\mz}{M_{Z}}

\newc{\OPALColl}   {OPAL Collaboration}
\newc{\ALEPHColl}  {ALEPH Collaboration}
\newc{\DELPHIColl} {DELPHI Collaboration}
\newc{\XLColl}     {L3 Collaboration}
\newc{\JADEColl}   {JADE Collaboration}
\newc{\CDFColl}    {CDF Collaboration}
\newc{\DXColl}     {D0 Collaboration}
\newc{\HXColl}     {H1 Collaboration}
\newc{\ZEUSColl}   {ZEUS Collaboration}
\newc{\LEPColl}    {LEP Collaboration}
\newc{\ATLASColl}  {ATLAS Collaboration}
\newc{\CMSColl}    {CMS Collaboration}
\newc{\UAColl}    {UA Collaboration}
\newc{\KAMLANDColl}{KamLAND Collaboration}
\newc{\IMBColl}    {IMB Collaboration}
\newc{\KAMIOColl}  {Kamiokande Collaboration}
\newc{\SKAMIOColl} {Super-Kamiokande Collaboration}
\newc{\SUDANTColl} {Soudan-2 Collaboration}
\newc{\MACROColl}  {MACRO Collaboration}
\newc{\GALLEXColl} {GALLEX Collaboration}
\newc{\GNOColl}    {GNO Collaboration}
\newc{\SAGEColl}  {SAGE Collaboration}
\newc{\SNOColl}  {SNO Collaboration}
\newc{\CHOOZColl}  {CHOOZ Collaboration}
\newc{\PDGColl}  {Particle Data Group Collaboration}

\def\issue(#1,#2,#3){{\bf #1}, #2 (#3)}
\def\iss(#1,#2,#3){{\bf #1} (#3) #2}
\def\ASTR(#1,#2,#3){Astropart.\ Phys. \issue(#1,#2,#3)}
\def\AJ(#1,#2,#3){Astrophysical.\ Jour. \issue(#1,#2,#3)}
\def\AJS(#1,#2,#3){Astrophys.\ J.\ Suppl. \issue(#1,#2,#3)}
\def\APP(#1,#2,#3){Acta.\ Phys.\ Pol. \issue(#1,#2,#3)}
\def\JCAP(#1,#2,#3){Journal\ XX. \issue(#1,#2,#3)} 
\def\SC(#1,#2,#3){Science \issue(#1,#2,#3)}
\def\PRD(#1,#2,#3){Phys.\ Rev.\ D \issue(#1,#2,#3)}
\def\PR(#1,#2,#3){Phys.\ Rev.\ \issue(#1,#2,#3)} 
\def\PRC(#1,#2,#3){Phys.\ Rev.\ C \issue(#1,#2,#3)}
\def\NPB(#1,#2,#3){Nucl.\ Phys.\ B \issue(#1,#2,#3)}
\def\NPPS(#1,#2,#3){Nucl.\ Phys.\ Proc. \ Suppl \issue(#1,#2,#3)}
\def\NJP(#1,#2,#3){New.\ J.\ Phys. \issue(#1,#2,#3)}
\def\JP(#1,#2,#3){J.\ Phys.\issue(#1,#2,#3)}
\def\PL(#1,#2,#3){Phys.\ Lett. \issue(#1,#2,#3)}
\def\ZP(#1,#2,#3){Z.\ Phys. \issue(#1,#2,#3)}
\def\ZPC(#1,#2,#3){Z.\ Phys.\ C  \issue(#1,#2,#3)}
\def\PREP(#1,#2,#3){Phys.\ Rep. \issue(#1,#2,#3)}
\def\PRL(#1,#2,#3){Phys.\ Rev.\ Lett. \issue(#1,#2,#3)}
\def\MPL(#1,#2,#3){Mod.\ Phys.\ Lett. \issue(#1,#2,#3)}
\def\RMP(#1,#2,#3){Rev.\ Mod.\ Phys. \issue(#1,#2,#3)}
\def\SJNP(#1,#2,#3){Sov.\ J.\ Nucl.\ Phys. \issue(#1,#2,#3)}
\def\CPC(#1,#2,#3){Comp.\ Phys.\ Comm. \issue(#1,#2,#3)}
\def\IJMPA(#1,#2,#3){Int.\ J.\ Mod. \ Phys.\ A \issue(#1,#2,#3)}
\def\MPLA(#1,#2,#3){Mod.\ Phys.\ Lett.\ A \issue(#1,#2,#3)}
\def\PTP(#1,#2,#3){Prog.\ Theor.\ Phys. \issue(#1,#2,#3)}
\def\RMP(#1,#2,#3){Rev.\ Mod.\ Phys. \issue(#1,#2,#3)}
\def\NIMA(#1,#2,#3){Nucl.\ Instrum.\ Methods \ A \issue(#1,#2,#3)}
\def\EPJC(#1,#2,#3){Eur.\ Phys.\ J.\ C \issue(#1,#2,#3)}
\def\RPP (#1,#2,#3){Rept.\ Prog.\ Phys. \issue(#1,#2,#3)}
\def\PPNP(#1,#2,#3){ Prog.\ Part.\ Nucl.\ Phys. \issue(#1,#2,#3)}
\newc{\PRDR}[3]{{Phys. Rev. D} {\bf #1}, Rapid  Communications, #2 (#3)}

\def\PLB(#1,#2,#3){Phys.\ Lett.\ B  \iss(#1,#2,#3)}
\def\JHEP(#1,#2,#3){JHEP \iss(#1,#2,#3)}

\vspace*{\fill}
\vspace{-1.5in}

\begin{center}
{\Large \bf Neutralino dark matter confronted by the LHC constraints on Electroweak SUSY signals }
  \vglue 0.4cm
  Arghya Choudhury\footnote{arghyac@iiserkol.ac.in} and
  Amitava Datta\footnote{adatta@iiserkol.ac.in}
      \vglue 0.1cm
          {\it 
	  Indian Institute of Science Education and Research - Kolkata, \\
          Mohanpur Campus, PO: BCKV Campus Main Office,\\
          Nadia, West Bengal - 741252, India.\\
	  }
	  \end{center}
	  \vspace{.1cm}

\vspace{+1cm}

\begin{abstract}
The supersymmetric particles (sparticles) belonging exclusively to the electroweak sector of the 
minimal supersymmetric standard model (MSSM) may hold the key to the observed dark matter relic 
density in the universe even if all strongly interacting sparticles are very heavy. The importance 
of the light EW sparticles in DM physics  and in producing spectacular collider 
signals is emphasized. It is shown that even the preliminary data on the direct searches of these 
sparticles at the LHC, significantly constrain the parameter space of the MSSM compatible with the 
observed relic density and provide useful hints about the future search prospects.  If in addition to the 
electroweak sparticles the gluinos are also within the reach of the LHC experiments, then  the gluino 
mass limits in the light slepton scenario obtained via the canonical jets + $\met$ channel may be relaxed by as 
much as 25 $\%$ compared to the existing limits. But the corresponding
same sign dilepton (SSD) + jets + $\met$ signal will yield enhanced limits competitive with the strongest 
limits currently available. This is illustrated with the help of benchmark scenarios at the 
generator level using PYTHIA. If the gluinos are just beyond 
the current reach of the LHC, then the generic n-lepton + m-jets + missing energy signal may discriminate 
between different DM producing mechanisms by comparing the signals corresponding to different values of n. 
This is illustrated by simulating the signals for n = 0 and n = 2 (the SSD signal).

\end{abstract}

\newpage
\setcounter{footnote}{0}
\section{Introduction}
Supersymmetry(SUSY) is a theory of elementary particles which for the first 
time relates bosons and fermions through symmetry transformations (see e.g., \cite{susy1,susy2,susy3}) . 
Apart from its aesthetic appeal this theory solves several practical problems. 
For example, the standard model (SM) of particle physics has no explanation of  
the observed dark matter (DM) relic density in the universe \cite{dmrev, dmrev1}. 
In contrast the minimal supersymmetric standard model (MSSM) 
\footnote{In this paper MSSM stands for a general model without specific assumptions regarding SUSY breaking.} 
with R-parity conservation provides an attractive DM candidate: the lightest neutralino 
($\lspone$)\cite{susy3,dm3}. This  weakly interacting massive particle (WIMP) is 
chosen to be the lightest supersymmetric particle (LSP) in many SUSY models.
By virtue of R-parity conservation it is stable and whenever an unstable
heavier superparticle (sparticle) is produced in an accelerator it eventually
decays into the LSP. 

Being a stable WIMP the LSP escapes detection, providing 
there by the missing energy signature - the hallmark of sparticle production. 
This is one of the many spectacular interconnections between particle 
physics and cosmology. Naturally the search for SUSY had been a top priority 
programme at the recently concluded experiments at the Large Hadron 
Collider (LHC) at 7 and 8 TeV \cite{atlas0l7,atlascms,atlas0l8,atlas2ssl8} and will continue to occupy the 
central stage of the upcoming LHC experiments at 13/14 TeV. 
These experiments mainly searched for strongly interacting sparticles (squarks and gluino) and 
non-observation of signals resulted in strong bounds on their masses.

A large number of phenomenological analyses have also addressed 
the issue of SUSY search at the LHC-7/8 TeV experiments \cite {susypheno}. 
The implications of the LHC constraints for neutralino DM in the MSSM \cite {dmmssm,boehm,r.catena, arghya3}
 as well as in models with specific SUSY breaking mechanisms \cite{dmsugra}have been studied in some 
recent analyses.

However, in many scenarios the most stringent bounds from the LHC and the DM
relic density are sensitive 
to different sectors of the MSSM parameter space. The LHC data may have little bearing on DM physics in such models. This intriguing possibility  was 
recently emphasized  in \cite{arghya3}.

The minimal supergravity (mSUGRA) model \cite{msugra}, also known as the constrained MSSM (cMSSM),
is a case in point. Here the above strong squark-gluino mass bounds can be easily translated into severe restrictions on the 
spectrum of the EW sparticles. As a result it is now well-known that the mSUGRA parameter space 
allowed by the relic density data has been severely depleted  by the LHC constraints \cite{dmsugra}.

In contrast the sparticle spectra in the strong and EW sectors are independent in the unconstrained MSSM. For a given strong sector consistent with the LHC data, there could be many
allowed EW sectors with different characteristics. 
This was illustrated in \cite{arghya3} with the help of several benchmark scenarios. 
These scenarios consist of different EW sectors each consistent with the relic density data.
It was found that the changes in the squark-gluino mass limits were rather modest (10 - 15 \%) of the order of the theoretical uncertainties 
in most scenarios 
\footnote{ In \cite{arghya3} as well as in this paper it is assumed that the masses of 
the sparticles in the strong and EW sectors are well separated. In compressed SUSY models \cite{martin} 
with approximate mass degeneracy between the above two sectors the squark-gluino mass limits 
are significantly weaker \cite{atlas0l7,compressed}}. 

In the unconstrained MSSM the EW sector is, therefore, restricted by the rather modest bounds from 
LEP \cite{lepsusy} and Tevatron \cite{tevatron} only. These bounds obtained by direct search of EW sparticles, 
are much smaller than the TeV landmark for supersymmetric models in general.
Model independent lower limit on the chargino mass from LEP data is 103.5 GeV \cite {lepsusy}.
The same on the slepton (smuon) mass is 96.6 GeV \footnote {This limit is valid for the R-type smuon.
However, it also yields a conservative limit for L-type sleptons having larger production cross section.} 
(for $\mlspone$ = 40 GeV). 
The limit from the unsuccessful trilepton search 
at the Tevatron is  145 GeV on chargino mass for a specific choice of parameters in the mSUGRA model. 

It was also illustrated in \cite{arghya3} that there are many purely electroweak annihilation and coannihilation 
mechanisms for DM production in the unconstrained MSSM which are not viable in the cMSSM (see Section 2.2 of \cite{arghya3}). 
Some examples are annihilation of a LSP pair into the lighter Higgs scalar resonance or 
the Z resonance, the LSP - sneutrino co-annihilation etc. 
Thus direct constraints on the EW sector from the LHC are of crucial importance.   

It is encouraging to note that the LHC data on direct production of EW gauginos 
(the charginos and the neutralinos) and charged sleptons \footnote {In this paper lepton (slepton) usually denotes 
a particle (sparticle) carrying $e$ or $\mu$ flavour. If the discussion involves $\tau$
or stau, it will be explicitly mentioned.}are now available (see below).
In this paper we shall extend  the analysis of \cite{arghya3} using more recent LHC data including the measurement of the Higgs boson mass \cite{higgs} 
and the results of the direct searches for the electroweak(EW)  sparticles.

Searches for direct production of electroweak sparticles have been carried out by both  
ATLAS \cite{atlas2l7,atlas3l7} and CMS \cite{cms2l3l7} using LHC data recorded at $\sqrt s$ = 7 TeV. Both ATLAS and CMS have 
updated these results at $\sqrt s$ = 8 TeV using 13.0 $\ifb$ 
\cite{atlas3l8} and  
9.2 $\ifb$ \cite{cms3l8} of data respectively.  
As no excess was observed in any of the channels studied so far, upper limits at 
95$\%$ CL were set on the relevant parameters in the R-parity conserving phenomenological MSSM (pMSSM) model 
\cite{pmssm} and in several simplified models.

These limits are presented in the figures of \cite {atlas2l7,atlas3l7,cms2l3l7,atlas3l8,cms3l8}.  
Some of the figures have been reproduced in this paper for 
ready reference (see the next section).
We wish to stress that these constraints are either independent of or very mildly
dependent on the strong sector.

Most of the above models contain  sleptons (either L or R-type) lighter than the lighter 
chargino and the second lightest neutralino. This feature spectacularly enhances the leptonic 
signals from the pair production of EW gauginos in general. Moreover, the current direct slepton searches 
at the LHC are already sensitive to light L-type sleptons.  

From the point of view of collider signals alone these models 
may appear to be only a subset of many possibilities in the MSSM. 
They, however, become more appealing in the context of neutralino DM.  
It is well known that 
the presence of the light sleptons provides 
efficient mechanisms for relic density production in the Universe. The importance of 
the R-sleptons in this respect has been  appreciated for a long time. 
It was recently emphasized in \cite{arghya3} that even with light L-sleptons and sneutrinos, 
the LSP-sneutrino coannihilation may turn out to be an important DM producing mechanism. 
Moreover, the contribution of light sleptons and gauginos can enhance the theoretical 
prediction  of the $(g-2)$ of the muon ($(g-2)_{\mu}$) which appears to be supported by 
experiment \cite{g-2exp}. This will be briefly discussed in a subsequent section.

If in addition to light EW sparticles, there are squarks and gluinos 
within the reach of 
future LHC experiments, 
the light slepton scenario may significantly modify the signatures of squarks- gluino production. 
Typically these signatures consist of final states with n - leptons + m - jets + $\met$, where 
n and m are integers. It is usually believed that the n = 0 case provides the best discovery 
channel or produces the strongest mass limits. In the light slepton scenario the EW gauginos 
present in the squark-gluino decay cascades, decay into final states 
involving e and $\mu$ with large branching ratios (BRs). This depletes a part of  
the zero lepton  signal weakening there by, the prospect of discovery or the mass limits. 
This was also demonstrated in the context of the Tevatron data on squark gluino 
production \cite{admgnp}. 
In contrast the signal with $n \geq 1$ will be correspondingly enhanced. 
On the other hand if the above gauginos dominantly decay into $\tau$ rich 
final states via a light stau, the 0l signal will be enhanced. These points were discussed \cite{arghya3} along with 
demonstrations in benchmark scenarios using 7 TeV data.

In this paper we shall address these issues in further details in the light gluino - heavy squark scenario 
using the data from the 8 TeV run. The possible enhancement of signals 
with $n \geq 1$ will be illustrated with the 
help of the same sign dilepton (SSD) + jets +$\met$ channel. We shall 
demonstrate that in some regions of the parameter spaces under 
consideration, the SSD channel may even provide better sensitivity 
compared to the 0l  channel.

The plan of the paper is  as follows. In Section 2 we shall consider several models 
constrained by the ATLAS and the CMS collaborations from electroweak SUSY signals 
and identify the parameter spaces compatible with both LHC and DM relic density data. 
We shall also consider some variants of the above models which leave the collider 
signals unaltered but provide elegant mechanism of relic density production. In Section 3
we shall discuss qualitatively some issues related to future electroweak SUSY signals. 
We shall also consider the modification of gluino mass limits in the light slepton scenario. 
The relative strength of the jets + $\met$ and SSD + jets + $\met$ signal will be advocated 
as the smoking gun signal of the light slepton scenario. 
We shall also briefly comment on $(g-2)_{\mu}$ in the light slepton scenario  and its compatibility with 
ATLAS \cite{atlas3l8} and the relic density data. Our main results will be summarized in the last section.

\section{ Neutralino DM in the MSSM with light EW sparticles}

We begin by briefly reviewing the chargino-neutralino sector of the MSSM. 
In the most general MSSM the charginos ($\chonipm $, i = 1,2) and the neutralinos 
($ \lspi $, i = 1-4) are  admixtures of the $SU(2)$ gauginos (the winos), 
the $U(1)$ gaugino (the bino) and the higgsinos 
(the superpartners of the Higgs bosons) with appropriate charges. 
These
mixings essentially depend  on  4 parameters - the gaugino mass parameters $M_1 $ and $M_2$, 
the Higgs mass parameter $\mu$ and tan$\beta$, 
the ratio of the vacuum expectation values of the two Higgs doublets. 
For  $|\mu|$ $>> |M_2| > |M_1|$, $\lspone$ is  bino ($\tilde B$) dominated and the lighter chargino  
$\chonepm$ (the second lightest neutralino $\lsptwo$) is mostly a charged (neutral) wino, but for $ |M_1| > |M_2|$, $\lspone$ 
($\lsptwo$) is dominantly the neutral wino (bino). On the other hand if $ |M_1| \simeq |M_2|$ the two lighter neutralinos are
admixtures of the neutral wino and bino.
In the limit, $|\mu|$ $<< |M_1|, |M_2|$, the two 
lighter neutralinos and the lighter chargino are all mostly higgsinos with approximately the same mass determined by $\mu$. 
If $|\mu|$ $\simeq |M_1| \simeq |M_2|$, some of the mass eigenstates will be strongly mixed. As we shall see below the ATLAS
and the WMAP data in conjunction shed light on the composition of the LSP in different scenarios. 

Through out this paper we have set $m_h \approx $ 125 GeV by taking the trilinear
soft breaking term $A_t$ as large (-3 to -4 TeV). No special SUSY breaking mechanism 
has been assumed. However, if any model considered in this paper can be motivated by a mSUGRA type scenario with radiative EW 
symmetry breaking and non-universal soft breaking terms,  we have pointed it out.

The ATLAS collaboration  have searched for chargino-neutralino ($\chonepm - \lsptwo$) pair
production \cite{atlas2l7,atlas3l7,atlas3l8} within the following frameworks. 

\begin{itemize}

\item The Light Gaugino and R-slepton (LGRS) Model: This is a phenomenological MSSM (pMSSM) \footnote{ The pMSSM is a model based on the MSSM with some extra assumptions like no flavour changing neutral current, no CP violation etc. It has 19 free parameters all defined at the EW scale \cite {pmssm}} inspired scenario with light R-type sleptons: 
Here tan $\beta =$ 6 and three representative values  $M_1$  (a) 100, (b) 140 and (c) 250 GeV have been considered. 
These models will be referred to as LGRSa, LGRSb and LGRSc respectively. The free parameters are $M_2$ and $\mu$. 
The masses of the R-type sleptons of all three generations are 
assumed to be degenerate and given by $ m_{\tilde l_R} = (\mlspone + \mlsptwo)/2 $. 
All other sparticles including L-type sleptons and sneutrinos are assumed to be
heavier than 2 TeV. In this case  $\chonepm $ decays exclusively to 
tau leptons as the chargino couples dominantly to the $\tau$ via its higgsino component.  
The $\lsptwo$ decays into charged lepton-antilepton pairs of all flavours with equal BRs. 

A WMAP allowed model sharing some features of the LGRS model considered in this paper and its signatures at 14 TeV LHC 
were recently discussed in \cite{boehm}.  However, the focus was on rather light bino like LSP 
with mass in the interval  1 GeV $\lsim \mlspone \lsim$ 30 GeV.  
Also much larger values of tan $\beta$ were assumed so that the lighter stau mass eigenstate was lighter than the R-type selectron or smuon. 
As a result LSP pair annihilation was dominated by stau exchange although other R-type sleptons also contributed. 
The chargino and the second lightest neutralino with significant Higgsino components,  decay dominantly  to the lighter stau. 
Thus the hadronically quiet trilepton signature looked for by the recent LHC experiments will be strongly suppressed in this case. However, the  
observability of tau rich final states arising from squark - gluino decay cascades in experiments at 14 TeV was discussed in \cite{boehm}.

\item The Light Gaugino and L-Slepton (LGLS) Model: In this simplified model the free parameters are 
the masses of $\chonepm, \lsptwo, \lspone, \tilde l_L,\tilde \nu $. It is assumed that the lightest neutralino 
is bino like and $\chonepm$ and $\lsptwo$  are wino like. All L-type slepton masses are 
chosen as $ m_{\tilde l_L} = m_{\tilde \nu} = (\mlspone + \mchonepm)/2 $. As a result the BR of chargino decay in  
slepton-neutrino and sneutrino-lepton mode of each flavour 
is the same. Similarly the $\lsptwo$ decay  into neutrino-sneutrino and lepton-slepton pairs of each flavour with equal probability.  

\item The Light Gaugino Heavy Slepton (LGHS) Model: This is a simplified model with heavy sleptons. 
In this scenario both L and R-type sleptons are assumed to be
heavy. Here $\chonepm$ and $\lsptwo$ decay only via on-shell or off-shell W and Z bosons respectively and the LSP. 
These heavier gaugino mass eigenstates are assumed to be wino dominated and mass degenerate while the
$\lspone$ is bino dominated. 

\end{itemize}

The CMS collaboration has also searched for final states with two, three and four leptons indicating 
the direct production of charginos and neutralinos \cite{cms2l3l7, cms3l8} in the LGLS, LGRS and LGHS models defined above. In contrast to the ATLAS analysis the compositions of the EW gauginos are fixed in this case.
The $\chonepm$ and $\lsptwo$ are assumed to be wino dominated and mass degenerate while the $\lspone$
is assumed to be bino like. 
However, both in LGLS and LGRS model the  
mass of the slepton ($m_{\tilde l}$) is allowed to vary and is parametrized by 
$m_{\tilde l}$ = $\mlspone$ $+$ $x_{\tilde l}$ $( \mchonepm - \mlspone)$, where 0 $<$ $x_{\tilde l}$ $<$ 1.

\begin{itemize}
\item We have also used the CMS constraints in the LGRS model with e) $x_{\tilde l}$ = 0.25, f) $x_{\tilde l}$ = 0.5 and 
g) $x_{\tilde l}$ = 0.75 which will be referred to as LGRSe, LGRSf and LGRSg respectively.  
\end{itemize}

Both the LHC collaborations have searched for slepton pair production via the opposite sign dilepton channel. 

\begin{itemize}
\item
In the the Light Left Slepton (LLS) Model it is assumed that only the L-sleptons and the sneutrinos are light and all
other sparticles are heavy. 
Thus the sneutrino is the 
the next lightest supersymmetric particle (NLSP) while the slepton decay into the LSP and a lepton with 100 \% BR. We shall use the ATLAS constraint
for a bino like LSP and have taken $\mu$ = 1000 GeV and tan$\beta$ = 6. 
\end{itemize}

The current LHC experiments are sensitive to L-type slepton pair production only. 
The cross-section of R-type slepton pair production is
too small to affect the present search results. However, they may have significant influence on DM production.

\begin{itemize}
\item We have also examined the Light Right Slepton (LRS) model constrained by the DM relic density alone. 
All sparticles other than the LSP and the R-type sleptons are assumed to be heavy. 
\end{itemize}

We shall also consider the following models with both R and L type sleptons having equal soft breaking mass parameters. In these model the possibility of L-R mixing among the third generation sleptons lead to interesting consequences in the context of both collider and DM physics.

\begin{itemize}
\item The Light Left and Right Slepton (LLRS) Model: 
We have considered two variants of the this model. 
First we have taken $\mu$ = 1 TeV as before. 
Here the $\mu$ tan$\beta$ term induces large mixing
in the stau mass matrix and the lighter stau mass eigenstate
($\stauone$)
could be significantly lighter than the other sleptons so that it
turns out to be the NLSP. This model will be referred to as the 
LLRS model with large mixing (LLRSLM).
    
Next we choose $\mu = 400.0$ GeV. Here all L and R
sleptons are approximately mass degenerate due to small mixing in the stau sector. As a result the sneutrino is still the NLSP. This is the LLRS model with small mixing (LLRSSM). 
\end{itemize}

In this paper we have computed the sparticle spectra and the decay BRs using SUSPECT (v2.41) \cite{suspect} and SDECAY \cite{sdecay}.
DM relic density has been computed by micrOMEGAs (v.2.4.1) \cite {micromegas}. 
The observed DM relic density ($\Omega h^2$) in the universe measured by the Wilkinson 
Microwave Anisotropy Probe (WMAP) collaboration \cite {wmap} is given by  $\Omega h^2$ = 0.1157 $\pm$ 0.0023. 
If 10$\%$ theoretical uncertainty is added \cite{baro} then the 
DM relic density is bounded by 0.09 $\leq \Omega h^2\leq$ 0.13 at 2$\sigma$ level. Recently the Planck collaboration 
has measured the same quantity and found  $\Omega h^2$ = 0.1199 $\pm$ 0.0027 at 68 $\%$ CL  \cite{planck}. 
This result is compatible with the WMAP data.

We now identify the parameter spaces allowed by  the LHC and the relic density constraints in the models discussed above :

\subsection {The LGRS Model : } 

Observed 95$\%$ CL exclusion contours in the $\mu - M_2 $ plane 
in the LGRSa, LGRSb, LGRSc models introduced above  have been obtained by 
ATLAS \cite{atlas3l7, atlas3l8} from unsuccessful chargino-neutralino searches. 
For ready reference we have presented them in Fig. 1(a), 1(b) and 1(c) for 
$M_1$ = 100, 140 and 250 GeV respectively. 
In each of these figures we have also superimposed the parameter space  
allowed by the observed DM relic density ( the shaded regions). 
The orange region is excluded by the LEP2 
constraint $\mchonepm >$ 103.5 GeV.

\begin{figure}[!htb]
\begin{center}
\subfloat[]{\includegraphics[angle =270, width=0.6\textwidth]{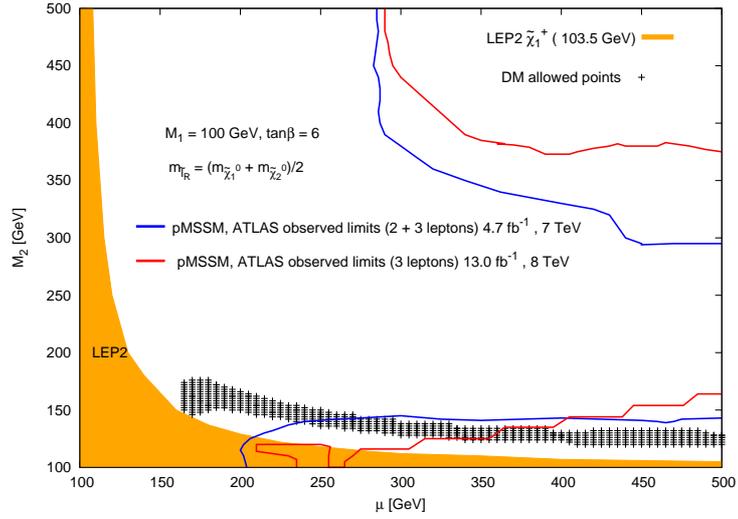} }\\
\subfloat[]{\includegraphics[angle =270, width=0.5\textwidth]{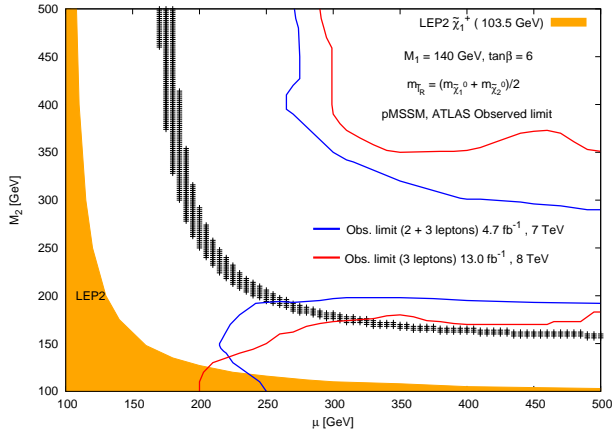} }
\subfloat[]{\includegraphics[angle =270, width=0.5\textwidth]{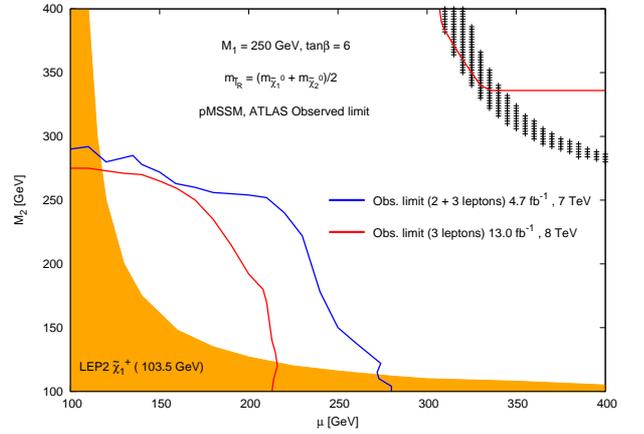} }
\caption{ {\footnotesize Red lines represent observed 95\% CL limit contours from chargino and neutralino production in 
the pMSSM by ATLAS \cite{atlas3l7,atlas3l8} for $M_{1}\,$$=\,$100$\,$GeV (a), $M_{1}\,$$=\,$140$\,$GeV (b) and $M_{1}\,$$=\,$250$\,$GeV (c). 
Blue lines represent the earlier limits from the combined 2l and 3l analysis \cite{atlas3l7} 
at the LHC 7 TeV run. Black (+ marked) points give 
WMAP allowed DM relic density.  }}
\end{center}
\end{figure}

In Fig. 1(a) the region between the upper red line and the lower red/blue line
are excluded by the ATLAS data. The parameter space  
for $\mu \simeq M_1$  and, in general, low values of $\mu$ are disfavoured by the ATLAS/LEP constraints 
irrespective of $M_2$. Thus a strongly mixed neutralino
(bino-wino-higgsino or bino-higgsino) is disfavoured. The limits disappear when both $\mu$ and 
$M_2$ are large (the upper right corner) as production cross-sections are smaller. 
However, this region is disfavoured by the relic density data.
For high  $\mu$  and low $M_2$  the mass differences between 
$\lspone$ and $\lsptwo$ or $\chonepm$ are rather small and there is no limit due to reduced sensitivity. 
This region is consistent with the observed DM relic density.  
In Fig. 1(a) the difference between $\mchonepm$ and $\mlspone$ in the region allowed by both LHC and WMAP experiments 
is typically 25 - 30 GeV.

Thus for relatively low $M_1$ both the ATLAS and WMAP data favour a Bino dominated LSP. 
In Fig. 1(a)  the LSP pair annihilation into $l^+l^-$ (bulk annihilation) and LSP - R-slepton 
co annihilation are the dominant relic density producing mechanisms. A modest fraction of the 
relic density is also produced by annihilation into the $W^+ W^-$ for relatively low $\mu$ 
which become rather tiny for $\mu \geq 500.0$. 

The same qualitative features are also observed in Fig. 1(b). 
In Fig. 1(b) the combination of ATLAS and WMAP constraints also favour a bino dominated LSP. 
However, a sizable wino and higgsino components are allowed for relatively small $\mu$. 
Here LSP annihilation into gauge boson pairs is the dominant source of relic density production 
while bulk annihilation as well as LSP -slepton coannihilation is also present. As $\mu$ increases 
the former  becomes less important and for $\mu \gsim 500$ it reduces 
to the 10 \% level.

In Fig. 1(c) with relatively large $M_1$  the situation is somewhat different. 
Here the parameter space between the blue line and the upper red line is allowed. 
The lower left corner corresponding to relatively low $\mu$ and $M_2$ is forbidden because of 
large cross sections. The upper right corner corresponding to relatively large $\mu$ and $M_2$ is 
disfavoured by the negative LHC search results since the mass difference between $\chonepm$ and the $\lspone$ 
is large. In the allowed region the mass difference between $\lspone$ and $\chonepm$ is small 
leading to lower sensitivity. 
It is interesting to note that the LHC constraints allow many different mixed and pure states 
for the LSP. However, in the WMAP allowed zone it is bino dominated with some wino and 
higgsino admixtures. As a result LSP pair annihilation into $W^+W^-$ and $t \bar{t}$ are the dominant 
DM producing mechanism with many other annihilation and coannihilation channels making small contributions. 

\begin{figure}[!htb]
\begin{center}
\includegraphics[angle =270, width= 0.9 \textwidth]{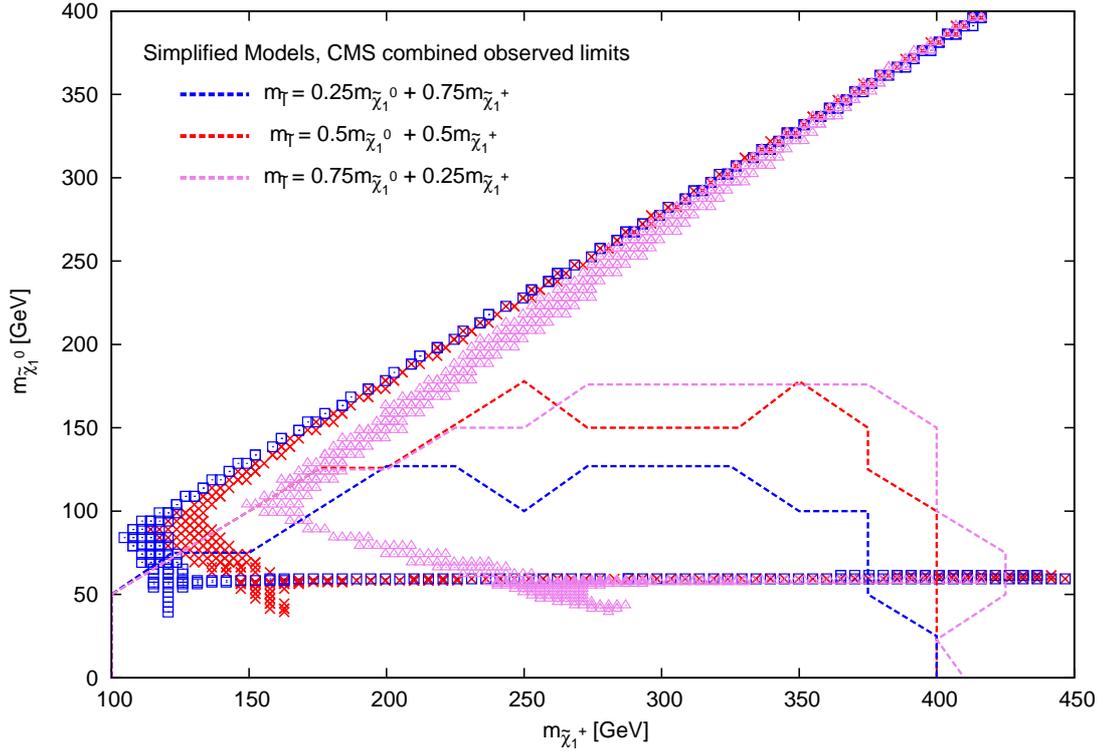}
\end{center}
\caption{ {\footnotesize  Observed 95 $\%$ CL limits in the simplified models with intermediate sleptons (R-type) decay 
by the CMS \cite{cms2l3l7}.  Here sleptons mass is chosen as $m_{\tilde l}$ = $\mlspone$ $+$ $x_{\tilde l}$ $( \mchonepm - \mlspone)$.
Violet, red and blue lines represent the exclusion contour for $x_{\tilde l}$ = 0.25, 0.50 and 0.75 respectively.  
Violet (triangle), red (x marked) and blue (square) points give WMAP allowed DM relic density 
following the same color convention. } }
\label{lgrscms}
\end{figure}

We have also considered the constraints  from CMS \cite{cms2l3l7} in the LGRS model with variable slepton mass. 
The 95 $\%$ CL exclusion contours are presented in Fig. \ref{lgrscms}. 
Exclusion contours for LGRSe, LGRSf, LGRSg models
are represented by  violet, red and blue lines. If the slepton mass, i.e., the parameter $x_{\tilde l}$, is varied, the limits on chargino - neutralino masses do change but not drastically.

Following the same color convention, WMAP allowed points  represented by 
triangles (violet), crosses (red) and squares (blue) are superimposed on Fig. \ref{lgrscms}. 
Here   $\lspone$ is bino like and $\chonepm$ is wino like. We have taken $\mu$ = 1000 GeV and tan$\beta$ = 6.

It follows from  Fig. \ref{lgrscms} that 
the observed DM relic density is obtained for $\mlspone$  $\approx$ $m_h /2$  due to  LSP pair
annihilation into the Higgs resonance. 
It may be noted that 
this LSP mass  is already ruled out for $\mchonepm <$ 370 GeV by the CMS search. For LSP masses little below $m_h/2$
the observed DM relic density may occur due to additional contributions from bulk annihilation via light R-sleptons.

Some parameter spaces adjacent to the $\mlspone \approx m_h / 2$ line  are also consistent with the observed
DM relic density for R-slepton masses in the small range  (100 - 110 GeV) just above the LEP limit (see Fig. \ref{lgrscms}). These regions  
correspond to different  $\mchonepm$ due to the fact that for smaller $x_{\tilde l}$,  
larger $\mchonepm$ yield the R-slepton mass in the above range.  
Here the main contribution to the relic density comes equally from three annihilation channels : 
$\lspone \lspone$ $\ra e\bar e$, $\mu \bar \mu$,  $\tau \bar \tau$. 
However bulk  of this parameter space is already disfavoured by the CMS data.

In Fig. \ref{lgrscms}, the R-slepton - LSP coannihilation strips corresponding to the observed DM relic density 
 are shown in the large neutralino mass region ($\mlspone > 100$ GeV). 
LSP pair annihilation into $e\bar e$, $\mu \bar \mu$ and $\tau \bar \tau$ final states still contribute 
60 to 80 $\%$ in different regions. However, $\tilde l_R$$\lspone$  coannihilations  into  final states with $\gamma l$ or $Zl$ vary between 
15 to 35 $\%$ depending on the parameter space. These WMAP allowed regions are well outside the present CMS exclusion 
contour.    

\subsection {The LGHS Model : }

\begin{figure}[!htb]
\begin{center}
\includegraphics[angle =270, width= 0.9 \textwidth]{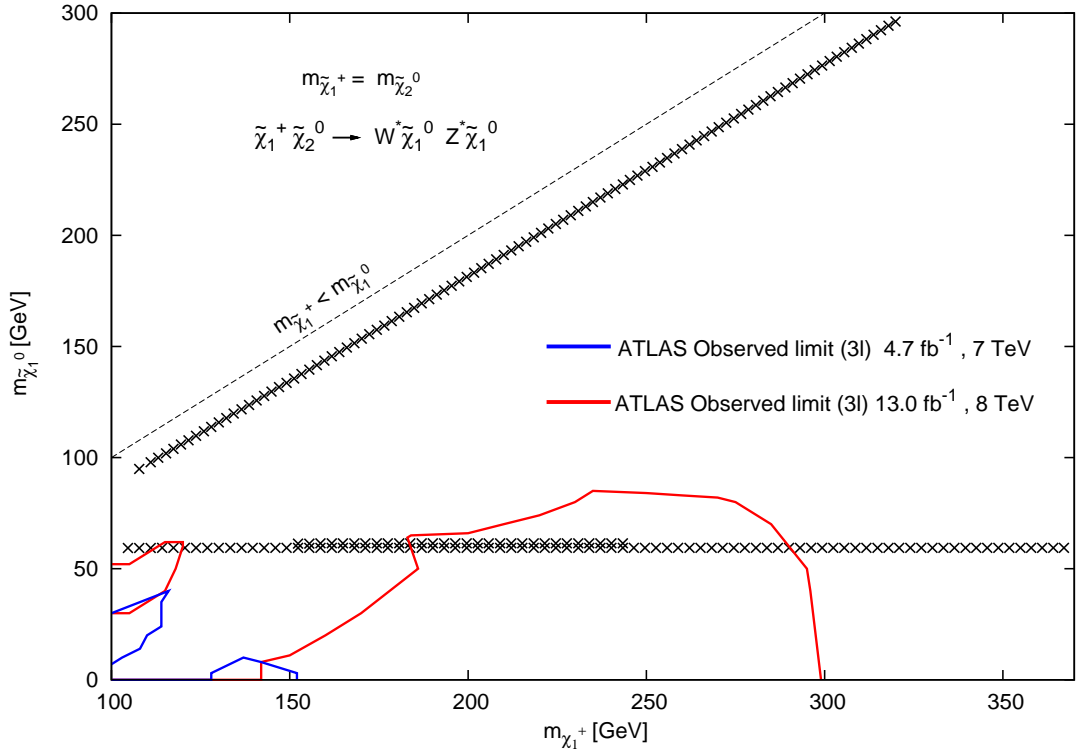}
\end{center}
\caption{ {\footnotesize  Red lines represent observed 95$\%$ CL limit contours for chargino and neutralino production in 
the simplified models (with heavy sleptons) by ATLAS \cite{atlas3l7,atlas3l8}. Blue lines correspond 
 to 7 TeV limits from 3l analysis \cite{atlas3l7}. The X marked black points are allowed by the WMAP data.  } }
\label{lghs}
\end{figure}

In Fig. \ref{lghs} the red lines represent the observed 95\% CL 
exclusion contours \cite{atlas3l8} in the simplified LGHS model 
where $\chonepm$ and $\lsptwo$ decay into the LSP 
and  gauge bosons - real or virtual. 
More stringent limits are obtained with the 8 TeV data compared to
the ones from the 7 TeV run \cite{atlas3l7} ( the blue lines). In this model $\charginopm$  masses upto 300 GeV 
are excluded for small $\mlspone$ and there is no limit for $\mlspone \gsim $ 90 GeV. The CMS limits \cite{cms3l8} are very similar.

The WMAP allowed parameter space represented by the black crosses has two
distinct regions. 
For $\mlspone$ $\simeq$ $m_h$/2, we get  WMAP allowed  DM relic density due to 
LSP pair annihilation into the Higgs resonance of mass $\simeq$ 125 GeV. 
The dominant relic density  producing mechanism is $\lspone \lspone$ $\ra h
\ra b\bar b$ (nearly 85 $\%$). Here the predicted relic density is 
to a large extent independent of other SUSY parameters. 
At present the collider data disfavours this LSP mass for a limited range of
chargino mass as shown in Fig. \ref{lghs} in this heavy slepton scenario. 

If the chargino - LSP mass difference is small we also get the correct DM relic density
along a coannihilation strip near the $\mchonepm$ = $\mlspone$ line. In this case several 
annihilation and co-annihilation processes among $\lspone$, $\lsptwo$ and $\chonepm$ contribute significantly.
We note in passing that  the ATLAS group has not considered the $\lsptwo$ decays via the $ h \lspone$ channel which could be important
in some regions of Fig. \ref{lghs}. For example, with  $\mchonepm$ = 300 GeV and 
$\mlspone \simeq$  60 GeV, the above mode is the most dominant decay channel of $\lsptwo $. This mode will deplete the 
trilepton  signal in the neighbourhood of the above point.   

The LGHS model can be motivated by a mSUGRA type model with radiative electroweak symmetry breaking and 
non-universal gaugino masses. Here the common scalar mass  $m_0$ and the $SU(3)_c$ gaugino mass $M_3$ are taken to be large at the GUT scale 
but $M_1$ and $M_2$ are relatively light. With $m_0$ = 2 TeV, $M_3$ = 1 TeV, $M_1$ = 300 GeV and $M_2$ = 180 GeV, we obtain the following 
light EW sparticles : $\mlspone$ = 124 GeV, $\mchonepm$ = $\mlsptwo$ = 138 GeV. 
Moreover from radiative EW symmetry breaking we obtain  $\mu$ = 2.1 TeV.
This ensures that $\lspone$ is bino dominated and $\lsptwo$, $\charginopm$ are wino dominated. 
With this spectrum we find $\Omega h^2$ = 0.10.

The LGHS model has also been constrained by indirect search for DM \cite{boehmpamela}. 
Such constraints -though interesting- involve uncertainties often associated with the interpretation of astrophysical data. 
This will be briefly discussed in Sec. 3.


\subsection {The LLS Model : } 

In Fig. \ref{slepton} we have presented the ATLAS exclusion contour (red lines) 
in the $\mlspone - \mslepl$ plane obtained from the search for  L-slepton (see footnote 4)
 pair production in the opposite sign dilepton channel. In this  
 model R sleptons and all EW gauginos except the LSP, are assumed to be heavy. 
Before the advent of the LHC, the LEP experiments had obtained a limit on the mass of the  $\tilde \mu_{R} $ as shown 
in Fig. \ref{slepton}. This limit implies that the left handed sleptons of equivalent masses are automatically excluded. 

\begin{figure}[!htb]
\begin{center}
\includegraphics[angle =270, width= 0.9 \textwidth]{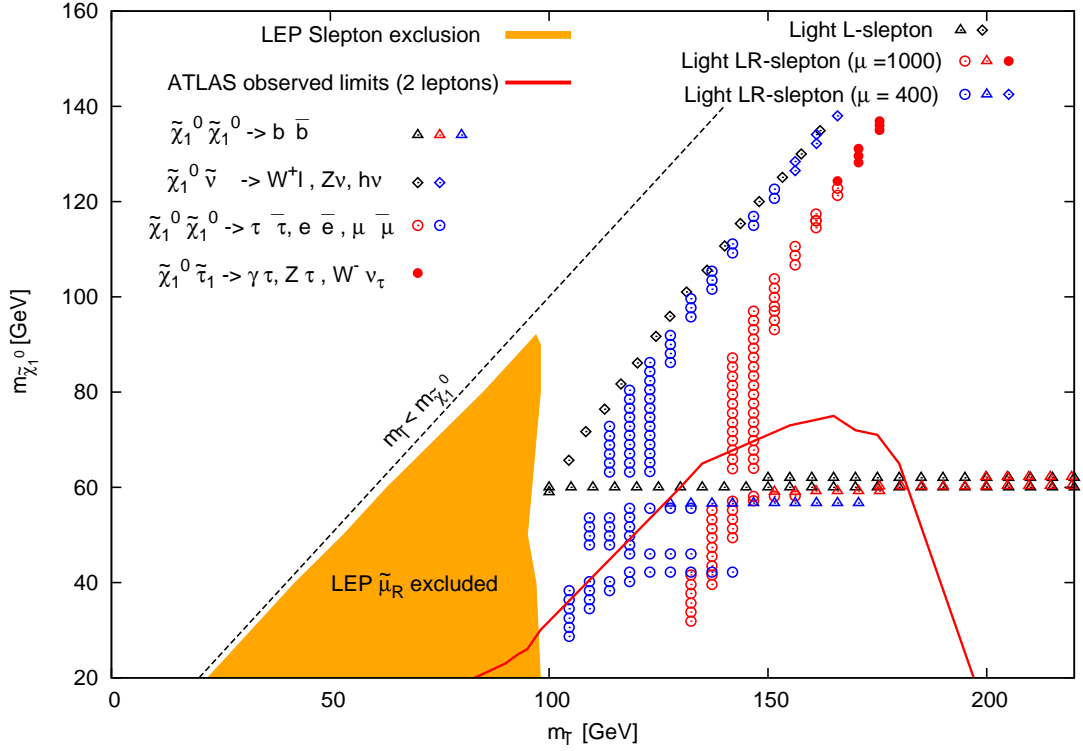}
\end{center}
\caption{ {\footnotesize The red line represents 95 $\%$ CL exclusion limits \cite{atlas2l7} for slepton pair production in 
the $m_{\tilde l}$ - $\mlspone$ plane in a pMSSM model. Yellow region is excluded by LEP experiments on
direct slepton pair production. Black, red and blue  colored points give WMAP allowed DM relic density in different models 
and the main channels contributing to the DM relic density are illustrated by points with different shape (see text for details). 
}}
\label{slepton}
\end{figure}

In order to calculate the DM relic density we have chosen $\mu$ =  1 TeV and tan $\beta$ = 6. 
With this choice the LSP is dominantly a bino. 
The points consistent with WMAP data are denoted by black motifs (triangle or diamond shape) for LLS scenario.  
As noted above the observed DM relic density may occur via LSP pair annihilation into the Higgs resonance 
(black triangle) for $\mlspone \approx m_h$/2 irrespective of the other sparticle masses. 
This LSP mass, however, is  already disfavoured for a range of slepton masses as shown in the Fig. \ref{slepton}. 

For larger LSP masses the region corresponding to appropriate relic density yields $\mlspone \simeq \msnu$. 
In this LLS scenario the sneutrino is the NLSP and  LSP - NLSP co-annihilation is the dominant DM producing mechanism for 
black diamond points. We have taken the standard formula involving the $SU(2)_L$ breaking D-term which 
relates the slepton and the sneutrino mass. This coannihilation strip (represented by diamond shape) is not yet sensitive to 
slepton searches as the  $\mslepl - \mlspone$ is rather small (25.0 GeV - 35.0 GeV ).  

\subsection {The LLRS Model :} 

It may be recalled that the presence of light R-sleptons does not affect the current LHC searches in this channel. 
These sleptons, however, significantly  affect the relic density production mechanism.

We have also indicated in Fig. \ref{slepton} the additional WMAP allowed regions for $\mslepl \simeq \mslepr$ and 
$\mu =$ 1000 GeV (LLRSLM model) by the red motifs (empty circle, triangle and filled circle). 
For relatively small  $\mlspone$ the LSP may dominantly pair annihilate into $\tau^+ - \tau^-$ 
to produce the DM relic density (see the vertical strips containing empty circles).  
Annihilation into other lepton-antilepton pairs also contributes albeit to a lesser extent. 
A sizable part of this region corresponding to low  LSP masses is already ruled out by the ATLAS data. 
However, for larger LSP masses a parameter space is still allowed by both WMAP and LHC data. Here 
LSP-stau coannihilation as well as a modest contribution from bulk annihilation produce the relic density.

This class of  model can be motivated by a mSUGRA type model with radiative electroweak symmetry breaking with non-universal gaugino masses if 
$m_0$, $M_1$, $M_2$ are taken to be relatively light while $M_3$ is large. 
With $M_1$ = 500 GeV, $M_2$ = 700 GeV, $M_3$ = 1 TeV and the common scalar mass $m_0$ = 190 GeV   
we obtained the following spectrum compatible with WMAP data : 
$\mlspone$ = 209 GeV, $\mlsptwo = \mcharginopm$ = 565 GeV, 
$\mstauone$ = 215 GeV, $\tilde {e_R}$ = 268 GeV, $\snutau$ = 471 GeV, 
$\snue$ = 482 GeV, $\tilde{e_L}$ = 488 GeV. Here $\mu$ = 1.9 TeV and  $\Omega h^2$ = 0.104. 

The points consistent with the WMAP data in the LLRSSM model are denoted by 
the blue motifs (circle, triangle and diamond shape). 
The pattern is roughly the same as the one exhibited by the red points.  
However, in the blue coannihilation strip corresponding to relatively large 
neutralino masses the sneutrino(NLSP) - LSP coannihilation is important. 
Bulk annihilation also contributes to some extent. It is clear that for
different choices of $\mu$ a large fraction of the parameter space
allowed by the ATLAS L-slepton search is consistent with the WMAP data.    

\begin{figure}[!htb]
\begin{center}
\includegraphics[angle =270, width= 0.9 \textwidth]{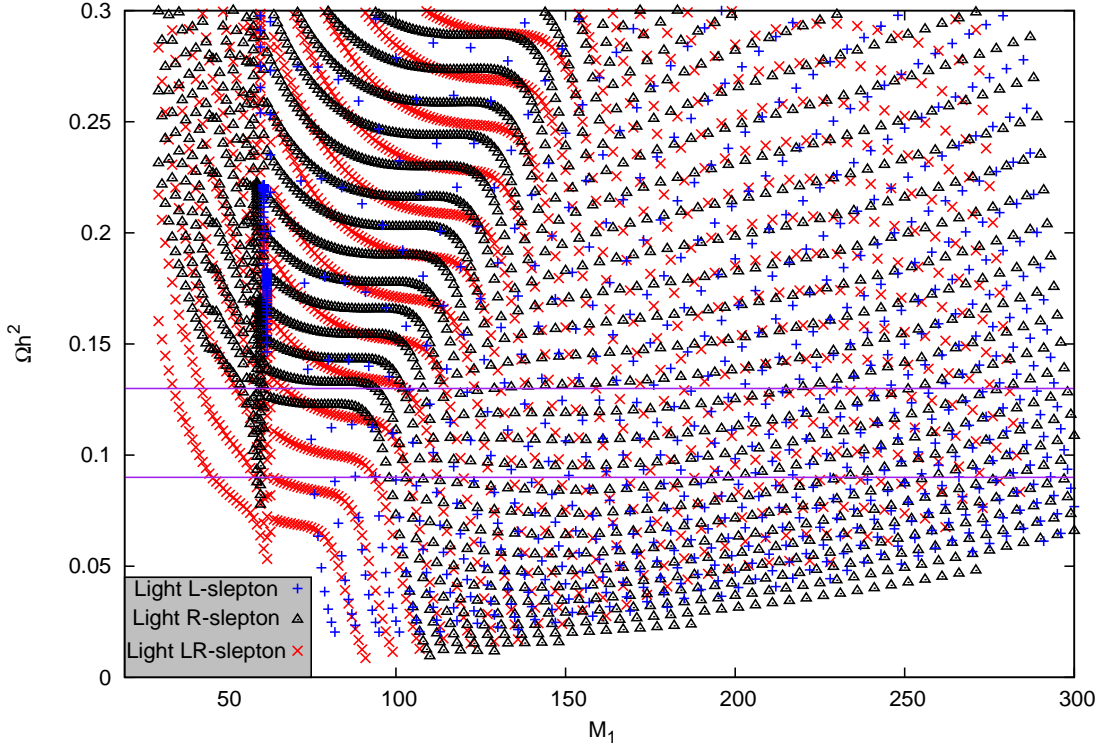}
\end{center}
\caption{ {\footnotesize $M_1$ vs $\Omega h^2$ for different choices slepton masses (see text for details). 
Purple horizontal lines represent the WMAP allowed DM relic density band (0.09 to 0.13).}  }
\label{slepton_arg}
\end{figure}

It follows from Fig. \ref{slepton} that rather small  ranges of LSP and slepton masses have been probed by  
the current slepton search experiments. We next examine the prospect of DM relic density production for larger LSP and slepton masses.
The WMAP allowed regions in the LLS, LRS and LLRSLM models are shown in Fig. \ref{slepton_arg} by the area between the two purple horizontal lines.  
In all models relic density production via the higgs resonance is allowed for $\mlspone \approx m_h / 2$. 
In each model we have varied the slepton mass input parameter (same for all three generations) in the range between 100 to 300 GeV
in steps of 5 GeV. We have taken $\mu$ = 1.5 TeV in each case and varied $M_1$  in the range between 0 to 300 GeV in steps of 1 GeV.

In the LLS model (blue + points of Fig. \ref{slepton_arg}) the DM producing mechanism is the same as in Fig. \ref{slepton}. 
In the LRS model (black triangles) bulk annihilation is the main DM producing mechanism 
for a small range of LSP mass just above the Higgs resonance provided the R-slepton 
mass is close to the LEP limit (see the near horizontal lines of Fig.  \ref{slepton_arg}). 
As the LSP mass increases LSP-R-slepton coannihilation represented by the sharply falling lines takes over and this
mechanism works for much larger LSP masses.

In the LLRSLM model (red cross marks) both left and right slepton mass input parameters are 
same and here the stau is the NLSP. The DM producing mechanisms are as in Fig. \ref{slepton} even for larger LSP masses. 

Some of the WMAP allowed points in Fig. \ref{slepton_arg} lead to interesting collider signals 
as we shall discuss in the next section.

\subsection {The LGLS Model : } 

The ATLAS collaboration has also searched for chargino-neutralino production in the LGLS model. The constraints
are summarized in Fig. 7(a) of \cite{atlas3l8}. We follow their choice
of the charged slepton mass: $ m_{\tilde l_L} = (\mlspone + \mchonepm)/2 $. However, we donot make the unrealistic choice   
$ m_{\tilde l_L} = m_{\tilde \nu}$. It may be noted that due to $SU(2)_L$
breaking D-terms the sneutrinos are always lighter than the corresponding 
charge sleptons. For computing the relic density and the collider signals we shall use 
the well known relation between slepton and sneutrino masses in the MSSM.

It is also worth noting that the modified sneutrino mass in the LGLS model may change in 
principle the bounds depicted in Fig. 7(a) of \cite{atlas3l8}. We have checked that for 
lighter sneutrinos the BR of the decay $\lsptwo \ra \snu \nu$ increases by a few percents. 
The sneutrinos decay invisibly via the channel $\nu \lspone$ . As a result the trilepton signal from $\chonepm - \lsptwo$ production is mildly suppressed. This weakens the limits obtained. 

The smaller neutrino mass facilitates the DM relic density production via sneutrino NLSP-LSP
co-annihilation. As in the LLS model this happens to be the most important relic density production mechanism. 

We shall conservatively use the somewhat stronger limits obtained in \cite{atlas3l8}. 
Although the chargino and the $\lsptwo$ are lighter in this model they are still heavier than the sleptons and contribute
negligibly to the relic density production. In the next section we shall choose benchmark points in this model 
taking into account the constraints shown 
in Figs. \ref{slepton} and \ref{slepton_arg} of this paper.          

\section{Collider Signals and Other Observables}

It follows from the last section that if $\mlspone \approx m_h/2$ the required DM relic density 
can be produced even if all other EW sparticles are heavy.
The monojet + $\met$ events \cite{monojet} can in principle probe this scenario. 
There are recent suggestions that this LSP mass can be probed by the upcoming direct search 
experiments \cite {r.catena,taohan,hooper}. 
This scenario reminds one of the importance of a $e^+e^-$ collider which can efficiently 
detect invisible particle production via the process $e^+e^-$ $\ra$ $\gamma + \met$. 
In this case the high energy photon  is generated by initial state radiation and can be detected easily in the 
clean environment of a  $e^+e^-$ collider \cite{photon1,photon2}. In the models where the  sneutrino is the NLSP, it decays invisibly. In such case this signal will be further enhanced \cite{photon1,photon3}. 

The scenario $\mlspone \approx m_h/2$ has already been constrained by the LHC data if  the LSP 
is accompanied by one or more relatively light EW sparticles. It can be readily seen from 
Figs 2 - 4 that $\mlspone \approx m_h/2$ is disfavoured for certain ranges of masses of the accompanying light sparticle/ sparticles.

Part of the WMAP allowed parameter space lying close to the existing exclusion contours in Figs. 1(a) - 1(c) is likely to be probed by 
the future experiments on the EW sparticles search. 
The same can be said about the bulk annihilation regions in the LLRSLM (red crosses) and LLRSSM (blue crosses) in Fig. \ref{slepton}.

However, the coannihilation strips in Figs. 2 - 4 will be difficult to probe at the LHC 
via the conventional electroweak signals. This is essentially due to the fact that they involve two or more 
nearly degenerate sparticles. 
It may be recalled that the LEP experiments obtained bound on a sparticle mass even if it is   
nearly degenerate with the LSP. 
It is, therefore, reasonable to believe that the sparticles producing the DM relic density via suitable 
co-annihilation mechanism can be successfully searched by the future $e^+e^-$ experiments.

There are scenarios, not discussed above, which can efficiently produce the DM relic density 
but the LHC signatures will be rather hard to detect in the near future. 
The light R-slepton (LRS) scenario is a case in point. Here the DM relic density can be efficiently produced 
by bulk annihilation and/or LSP - R-slepton coannihilation. However, 
as already noted the R-slepton pair production cross section at the LHC  is too small. 
Again a $e^+e^-$ collider will be best machine for probing this scenario.

\begin{table}[!htb]
\begin{center}
\begin{tabular}{|c|c|c|c|c|c|c|c|c|}
\hline
Benchmark  &Taken	&$\mlspone$	&$\mchonepm$	&$\mlsptwo$	&$m_{\wt l_L}$	&$m_{\wt l_R}$	&$\mstauone$	& $m_{\wt \nu_L} $	\\
Points	   &From	&		&		&		&($l=e,\mu$)	&($l=e,\mu$)	& 		&			\\
	   &		&(GeV)		&(GeV)		&(GeV)		&(GeV)		&(GeV)		& (GeV)		&(GeV)			\\
\hline
BP1	 &Fig. 1(a)	&97		&127		&127		&2000		&112		&112 		&2000			\\
\hline
BP2	 &Fig. 1(c)	&238		&271		&276		&2000		&257		&257 		&2000			\\
\hline
BP3	 &Fig. 3	&62		&152		&152		&2000		&2000		& 2000		&2000			\\
\hline
BP4	 &Fig. 3	&62		&351		&351		&2000		&2000		& 2000		&2000			\\
\hline
BP5	 &Fig. 3	&150		&166		&166		&2000		&2000		& 2000		&2000			\\
\hline
BP6	 &Fig. 2	&149		&206		&206		&2000		&163		& 163		&2000			\\
\hline
BP7	 &Fig. 4	&96		&610		&610		&128		&2000		&128 		&103			\\
\hline
BP8	 &Fig. 5	&199		&257		&257		&224		&2000		&224 		&211			\\
\hline
BP9	 &Fig. 5	&292		&375		&375		&313		&2000		&313 		&304			\\
\hline
BP10	 &Fig. 5	&240		&341		&341		&273		&273		&244 		&263			\\
\hline

       \end{tabular}
       \end{center}
          \caption{ {\small The mass spectra corresponding to different benchmark points
taken from different regions  of Fig. 1 - Fig. 5 consistent with the LHC and WMAP constraints. }}
\label{tab1}
          \end{table}

\begin{table}[!htb]
\begin{center}
\hspace{-1 cm}
\begin{tabular}{|c|c|c|c|c|c|c|c|c|c|c|c|c|c|c|}
\hline
Decay 					&BP1 	&BP2	&BP3	&BP4	&BP5	 &BP6	&BP7	&BP8	 &BP9	&BP10	\\
Modes					&	&	&	&	&	 &	&	&	 &	&	\\
\hline

$\gl    \ra \lspone  q \bar q  $			&11	&11	&11	&17	&11	 &12	&60	&12	 &15	&15	\\
$\quad  \ra  \chonepm q q\prime$	&55	&43	&59	&56	&59	 &58	&26	&58	 &56	&56	\\
$\quad  \ra  \lsptwo q \bar q  $		&28	&20	&30	&27	&30	 &30	&14	&30	 &29	&29	\\

\hline
\hline

$\chonepm \ra \lspone q q\prime $	&-	&-	&-	&-	&67	 &-	&-	&-	 &-	&-	\\
$\quad    \ra \lspone \ell \nu_{\ell} $	&-	&-	&-	&-	&33	 &-	&-	&-	 &-	&-	\\
$\quad    \ra \lspone W        $	&-	&-	&100	&100	&-	 &-	&-	&-	 &-	&-	\\
$\quad    \ra\snutau \tau	$	&-	&-	&-	&-	&-	 &-	&17	&22	 &19	&19	\\
$\quad    \ra\stau_{1} \nutau  $	&100	&100	&-	&-	&-	 &100	&16	&12	 &14	&13	\\
$\quad    \ra\snul l	$		&-	&-	&-	&-	&-	 &-	&34	&43	 &37	&37	\\
$\quad    \ra \wt{l}_L \nu_{l}$		&-	&-	&-	&-	&-	 &-	&32	&23	 &29	&29	\\

\hline\hline
$\lsptwo  \ra \lspone h 	     $	&-	&-	&-	&92	&-	 &-	&-	&-	 &-	&-	\\
$\quad  \ra  \lspone Z 	     	    $	&-	&-	&-	&8	&-	 &-	&-	&-	 &-	&-	\\
$\quad  \ra  \lspone \gamma 	    $	&-	&-	&-	&-	&25	 &-	&-	&-	 &-	&-	\\
$\quad  \ra  \slepl^\pm l^\mp       $	&-	&-	&-	&-	&-	 &-	&33	&25	 &30	&29	\\
$\quad  \ra\snul \bar{\nu_{l}}      $	&-	&-	&-	&-	&-	 &-	&33	&42	 &37	&36	\\
$\quad  \ra  \slepr^\pm l^\mp       $	&67	&67	&-	&-	&-	 &65	&-	&-	 &-	&-	\\
$\quad  \ra  \stauonepm \tau^\mp    $	&33	&33	&-	&-	&-	 &35	&17	&12	 &15	&14	\\
$\quad  \ra  \snutau \bar\nutau	$	&-	&-	&-	&-	&-	 &-	&17	&21	 &18	&18	\\
$\quad  \ra  \lspone q \bar q       $	&-	&-	&70	&-	&60	 &-	&-	&-	 &-	&-	\\
$\quad  \ra  \lspone\ell^\pm \ell^\mp$	&-	&-	&10	&-	&5	 &-	&-	&-	 &-	&-	\\
$\quad   \ra  \lspone\nu \bar \nu    $	&-	&-	&20	&-	&10	 &-	&-	&-	 &-	&-	\\
\hline
       \end{tabular}
       \end{center}
          \caption{ {\small The BRs ($\%$) of the dominant decay modes of $\gl$ (for $\mgl$ = 1 TeV),
 $\chonepm$ and $\lsptwo$  for the benchmark points. Here $l$ stands for e and $\mu$, but $\ell$ denotes all three generation 
leptons. } }
\label{tab2}
          \end{table}

If one or more strongly interacting sparticles are just beyond the current reach of the LHC 
experiments, then some of the light sleptons scenarios allowed by the WMAP data may lead 
to distinctive signatures in the n-leptons + m-jets + $\met$ channel for different values of n. 
This was demonstrated by the light stop scenario or the light stop-gluino (LSG) scenario 
in \cite {arghya3}. 
Several other groups have also considered such possibilities \cite{lightstop}. 

In this paper we shall concentrate on the light gluino - heavy squark scenario. 
We substantiate the above claim with the help of several benchmark points  
chosen from the models introduced in the last section. 
All points are allowed by LHC and WMAP constraints. 
In the following discussion we take $\mgl$ = 1 TeV 
which is approximately the current gluino mass limit if all squarks are heavy. 

The mass spectra corresponding to our benchmark points are presented in Table \ref{tab1}. The decay modes relevant for the gluino 
signal and their branching ratio (BRs) are included in Table \ref{tab2}. Next we will discuss the 
characteristics of these points. 

BP1 and BP2 are taken from LGRSa and LGRSc scenarios (Figs. 1(a) and 1(c)) respectively. 
Note that the  decay modes and BRs are very similar for these two points. However, 
the gaugino masses are relatively large in BP2. For reasons already discussed the 
 $\charginopm$ dominantly decays to $\stau_{1} \nutau$.

BP3, BP4, BP5 are chosen from the LGHS scenario (Fig. \ref{lghs}). 
In BP3 and BP4,  $\mlspone$  $\approx$ $m_h/2$. 
In both the cases the correct DM relic density 
is produced due to the Higgs resonance. 
In BP4, $\mcharginopm (\approx \mlsptwo$) is much larger than that in BP3. As a result 
$\lsptwo$  dominantly decays into $h \lspone$ which suppresses the leptonic signals. 
For BP5 lying on the coannihilation strip in Fig. \ref{lghs}, $\mcharginopm$ ( = $\mlsptwo $)
and $\mlspone$ are very close. This also suppresses the leptonic signals as we shall see below.

BP6 is chosen from LGRSe model (Fig. \ref{lgrscms}) where right handed slepton mass = 
0.75 $\times \mlspone$  + 0.25 $\times \mchonepm $ ( $x_{\tilde l}$ = 0.25). 
Here the LSP-slepton mass difference is rather small while chargino-slepton mass difference is large. 
In  this case the relevant decay modes and their BRs are similar 
to BP1 or BP2. We have also considered $x_{\tilde l}$ = 0.25 and 0.5, but the gluino signal discussed below 
does not show any new feature. 

BP7 is chosen from LLS scenario (Fig. \ref{slepton}). 
The chargino is much heavier than the slepton but lighter than the gluino. 
This does not affect the slepton pair 
production signal.  Here  BR ($\gl    \ra \lspone  q \bar q $)  is 60 $\%$. 
As a result the SSD signal will be suppressed (see Table \ref{tab3}). This point is also allowed by the 
chargino - neutralino search in LGLS model (see Fig. 7(a) of \cite{atlas3l8}). 

BP8, BP9  belong to the LGLS model discussed at the end of the last section. These points  are 
allowed by the chargino - neutralino search (see Fig. 7(a) of \cite{atlas3l8}) and  consistent with  the WMAP data (see Fig. \ref{slepton_arg}).

The point BP10 in the LGLRS scenario with both light L and R type sleptons is consistent with the WMAP data (Fig. \ref{slepton_arg}) 
and all LHC search results.

Next we turn our attention to the n-leptons + m-jets + $\met$ (n $\ge$ 0) signal in the light gluino scenario. 
We have concentrated on the gluino mass limits obtained in the n = 0 and n = 2 (SSD) channel using the current ATLAS data. 
 
ATLAS group has updated their result for SUSY search in jets + $\met$ channel  (n = 0) for $\lum$ = 5.8 $\ifb$ 
at 8 TeV \cite {atlas0l8}. They have defined five inclusive analysis channels labelled as A to E 
according to jet multiplicity from two to six. The details of the cuts are given in Table 1 of \cite {atlas0l8}. 
Depending upon the final cuts on the observables $\met$ / $m_{eff}$ and $m_{eff}$(incl.)
each channel is further classified as `Tight',`Medium' and `Loose'. 
ATLAS has finally presented the results for 12 signal regions and has constrained 
any new physics model in terms of upper limit on the effective cross-section $\sigma_{BSM}$/fb 
or number of events $N_{BSM}$.  
The observed upper limits on $N_{BSM}$ at 95 $\%$ Confidence Level (CL) for 
signal regions SRA-Tight, SRA-Medium, SRA-Loose, SRB-Tight, SRB-Medium, SRC-Tight, 
SRC-Medium, SRC-Loose, SRD-Tight, SRE-Tight, SRE-Medium, SRE-Loose are 
8.9, 33.9, 224.8, 7.3, 43.8, 3.3, 17.9, 65.7, 6.0, 9.3, 9.9, 10.4 respectively \cite{atlas0l8}. 

Event selection criteria and details of the cuts for two same sign dilepton search by ATLAS 
are available in \cite {atlas2ssl8}. The observed upper limits on $N_{BSM}$ at 95 $\%$ confidence level (CL) 
for this channel at $\lum$ = 5.8 $\ifb$ is 6.3. 

We have adopted different selection criteria for different signal regions  and checked whether 
the no of events coming from $\gl \gl$ pair production exceed the corresponding upper limit for at least 
one signal region.  In this way we have derived the new limits. 
Using PYTHIA (v6.424) \cite{pythia} Monte Carlo (MC) event generator, we have generated
the signal. The next to leading order (NLO) cross-section for the $\gl \gl$ pair 
production have been computed by PROSPINO 2.144 \cite {prospino} with CTEQ6.6M PDF \cite {cteq6.6}.

Limits on $\mgl$ using the ATLAS jets + $\met$ \cite {atlas0l8} data and the SSD \cite {atlas2ssl8} data 
are presented in Table \ref{tab3}. In most cases the SRD-Tight signal region \cite {atlas0l8} is effective in finding the revised exclusion limit for different EW sectors.

It is clear from Table \ref{tab3} that in the LGHS model strongest limits ($\mgl \ge$ 1150 GeV) come from 
ATLAS jets + $\met$ \cite {atlas0l8} data for BP3 - BP5. 
On the other hand the SSD signal is severely depleted in BP4 ( due to the spoiler mode $\lsptwo$ $ \ra h \lspone$ ) 
and BP5 (due to small mass differences among the EW gauginos). 
It will be interesting to check whether the higgs boson in the final state can be reconstructed to distinguish 
BP4 from other scenarios. 
In BP5 a fair fraction of the final state will contain an isolated photon (see Table \ref{tab2}). 
In fact we have checked that all along the coannihilation strip in Fig. \ref{lghs} the decay  $\lsptwo$ $ \ra \gamma \lspone$ 
occur with moderately large BR.
It will be challenging to nail down this scenario by identifying 
the photon.

\begin{table}[!htb]
\begin{center}\
\begin{tabular}{||c||c||c||}
\hline
Points		& \multicolumn{2}{c|}{Limit on $\mgl$ (GeV)} 	\\
\cline{2-3}
		& $Jets+ \met$ data \cite{atlas0l8}	& SSD data\cite{atlas2ssl8}		\\
\hline
BP1	 	& 	1050 		&	850 		\\
\hline
BP2	 	&	950 		& 	760		\\
\hline
BP3	 	&	1130 		& 	740		\\
\hline
BP4	 	& 	1150 		&	- 		\\
\hline
BP5	 	&	1140 		& 	-		\\
\hline
BP6	 	&	1050 		&  	840		\\
\hline
BP7	 	& 	1010 		&	- 		\\
\hline
BP8	 	& 	880 		&	960 		\\
\hline
BP9	 	&	730 		& 	930		\\
\hline
BP10	 	&	780 		& 	960		\\
\hline

       \end{tabular}\
       \end{center}
           \caption{Limits on $\mgl$ using the ATLAS jets + $\met$ \cite {atlas0l8} data and the SSD \cite {atlas2ssl8} data. }
\label{tab3}
          \end{table}


For BP1, BP2, BP6 and BP7 gluino mass limit is about 1 TeV from the jets + $\met$ data. 
The SSD channel puts weaker constraints. 
In BP7 with relatively heavy chargino, the limits from the SSD channel disappears.

In BP8, BP9, BP10 jets + $\met$ data puts a weaker limit on $\mgl$ (see Table \ref{tab3} for details). 
In fact in all cases the SSD (2l) signal is more sensitive to $\mgl$ than the $0l$ signal. 
It may be noted that all these points are either from LGLS or LGLRS scenario i.e., the models contain light L-sleptons.

The above discussions suggest that should a SUSY signal be seen, the relative size of 0l and SSD signal 
may provide some hint for the underlying electroweak sector with light sleptons. 
In Table \ref{tab4} we present the ratio $r_1$ of  0l events (with SRD-Tight signal region) \cite {atlas0l8} 
and the SSD events \cite {atlas2ssl8} for $\mgl$ = 1.2 TeV. 
In the second column the sparticle spectra as in Table \ref{tab1} have been used to compute $r_1$. In the third column only
the light sleptons in each scenario are replaced by heavy sleptons for computing the same ratio denoted by $r_2$. 
It may be noted that irrespective of the electroweak sector concerned $r_1 < 10$. 
On the other hand in other scenarios not presented in Table \ref{tab4}, $r_1$ is very large. This ratio, free from many theoretical uncertainties, 
may turn out to be an useful model discriminator.

\begin{table}[!htb]
\begin{center}\
\begin{tabular}{||c||c||c||}
\hline

Points		& With light slepton	& Without light slepton		\\
		& $r_1$ = $\frac{S(0l+j+ \etslash)} {S(2l+j+ \etslash)}$& $r_2$ = $\frac{ S(0l+j+ \etslash)} {S(2l+j+ \etslash)}$\\
\hline
BP1	 	& 	5.04 		&	19.52 		\\
\hline
BP2	 	&	3.89 		& 	12.16		\\
\hline
BP6	 	&	7.99		& 	35.36		\\
\hline
BP7	 	& 	 2.59		&	56.28		\\
\hline
BP8	 	& 	 1.81		&	39.65		\\
\hline
BP9	 	& 	1.51 		&	28.17		\\
\hline
BP10	 	& 	 1.49		&	28.33		\\
\hline

       \end{tabular}\
       \end{center}
          \caption{Here $r_1$ ( $r_2$ ) represents the ratio of  0l signal, with SRD-Tight signal region \cite {atlas0l8}, 
and 2l signal \cite {atlas2ssl8} for $\mgl$ = 1.2 TeV for light (heavy) sleptons.}
\label{tab4}
          \end{table}

Interest in the light slepton scenario builds up in the context of the 
discrepancy  between the measured value of the (g-2) of the muon \cite{g-2exp} and the standard model prediction. 
For a review of the theoretical prediction see, e.g., \cite{muonrev}.
The  alleged disagreement,however, hinges on the belief that 
the tiny corrections to $ (g-2)_{\mu}$ due to low energy strong interactions can be computed very accurately. 
It is claimed  that this can be achieved by computations using the $e^+e^-\ra$ hadron data.  
One then finds that a positive contribution from physics beyond the SM of 
$ \Delta a_\mu \equiv a_\mu({\rm exp}) - a_\mu({\rm SM})= (26.1 \pm 8.0) \times 10^{-10} $ 
is required to  
resolve this discrepancy. It is worth noting that if the same corrections are computed using 
the hadronic $\tau$ decay data the agreement between the SM prediction and the measured value improves. 
For a recent critical appraisal of different methods of computing the hadronic contribution and possible sources of error
see, e.g., \cite{sirlin}.

It is well-known that SUSY  can potentially enhance the $(g-2)_{\mu}$. Moreover, it can be computed in the models considered
in this paper in a straightforward way, without invoking additional assumptions. 
The major SUSY contributions may arise from i) the chargino - sneutrino loop or ii) the neutralino - smuon 
(both L and R - type) loop \cite{susyg-2}. Moreover, these contribution can be computed in the models described in the previous section without invoking any new assumptions. 

SUSY contributions of type i) naturally arises in the LGLS model. Recently this model has been 
constrained by simultaneously using the ATLAS data on chargino - neutralino search and the $(g-2)_{\mu}$ data 
(see Figs. 1(a) - 1(c) of \cite{endo})  These constraints can not be directly compared with the similar ones obtained by ATLAS \cite{atlas3l8},  
as they were obtained with different MSSM inputs. However, the WMAP allowed regions in 
Figs. 1(a) - 1(b) of  \cite{endo} can be easily identified. It has already been noted in the last section that the 
relic density in this model is produced mainly by sneutrino-LSP co-annihilation. 
Thus the WMAP allowed regions must lie just above the thick dark lines, representing 
$\msnu = \mlspone$ shown in these figures. Using micrOMEGAs \cite {micromegas}  we have checked 
that these points simultaneously satisfy the LHC, the DM relic density and the $(g-2)_{\mu}$ constraints.

The contribution of type ii) arises in the light LGLRLM
model with large  higssino
mass parameter $\mu$. The EW signals in this model
have not been investigated by the LHC experiments. However,
as in the LGLS model the relevant constraints have been presented in Fig. 1(d) of \cite{endo}. 
In the model of \cite{endo} the right sleptons
are heavier than the L-sleptons and the sneutrino is the NLSP. As a result
the sneutrino is the NLSP and NLSP - LSP coannihilation is the dominant DM producing mechanism.
Thus
the region just above the thick black line of Fig. 1(d) of \cite{endo} represent
the parameter space consistent with the WMAP data which is also allowed by
the $(g-2)_{\mu}$ and LHC constraints. 

Several direct \cite{cdms, xenon100} DM search experiments have been performed recently. Interesting constraints on  neutralino DM have been reported. As an example let us consider the XENON100 experiment \cite {xenon100} which
obtained an upper bound on the spin independendent DM-nucleon scattering cross-section ( $\sigma_P^{SI}$ )as a function of
the mass of the proposed DM particle. Following the standard procedures one can compute this cross section in a given model of neutralino DM and restrict the parameter space using the above upper bounds. 

As an illustration one can look into Fig 5 of \cite{taohan}. It follows that although some parameter 
space is excluded by XENON100 experiment, the entire range of neutralino mass considered by us and many 
DM producing mechanisms corresponding to these masses are still allowed. Using representative  points we 
have also checked that the parameter space probed by us are by and large consistent with the 
XENON100 constraints. For example, in the LLRSSM model (Fig. \ref{slepton}) with $\mu = 400$ GeV 
the cross section  varies between 4.7 $\times$ $10^{-10}$ pb and 1.7 $\times$ $10^{-9}$ pb for $ 40 \lsim \mlspone \lsim 160$ GeV. 
In the LLS and LLRSLM models with $\mu = 1000$ GeV  the cross section  varies between 
5.9 $\times$ $10^{-11}$ pb and 1.1 $\times$ $10^{-10}$ pb for the same LSP mass range. The above
cross sections lie well below the corresponding XENON100 upper limits.             

It should be borne in mind that the computation of $\sigma_P^{SI}$ involve large
uncertainties. For example, the DM relic density near the earth could be significantly different from the usually assumed average value, due to clumpiness in the DM distribution. This may significantly change the neutralino flux and , hence, the above upper
bound.  More important, the above  cross section depends on nuclear form factors not precisely known. Recently it has been noted that a new determination of the strangeness content of the nucleon  further relaxes the upper bound on $\sigma_P^{SI}$ which allows more parameter space. Some of 
these uncertainties have been discussed in the latest version of micrOMEGAs 3.1 (see \cite{micromega3} and references therein). This clearly demonstrates that the conclusions from the XENON experiments are yet to  stabilize.

Since DM is distributed all over the universe, it natural to expect that 
various astrophysical data can potentially shed light on DM physics.
These experiments, popularly known as indirect detection experiments,
have reported some anomalies in the measurement of some cosmic ray fluxes  \cite{pamela,fermilat}. However, opinion about there results differ 
(for a list of critiques, see, e.g., refs 6 -12 of \cite{boehmpamela}).
 
We now briefly comment on a few searches which are less controversial. 
A  discussion on the corresponding uncertainties can be found , e.g., in \cite{boehmpamela}. 
The Fermi-LAT collaboration has examined diffuse $\gamma$-ray emission in dwarf spheroidal (dSph) galaxies \cite{fermilatdsph} 
and also in the Milky Way \cite{lat1,lat2}. Dark Matter candidates with a total annihilation cross section of 
$\bra \sigma v \ket$ = 3 $\times 10^{−26}$ cm$^3/s$ have been ruled out if the mass of the DM candidate is 
$\lsim $ 30 GeV, although the limit weakens considerably for higher DM masses. Moreover the 
analysis of ref \cite{fermilatdsph} indicates that the uncertainties associated with the DM energy density profile of individual
dSph galaxies may relax the bound significantly. 

In ref \cite{boehmpamela} bounds on the pair annihilation cross section of a generic DM candidate 
into $W^+ W^-$ was obtained as a function of DM mass using the PAMELA measurement of 
the galactic antiproton flux \cite{pamela2,pamela3}. The results were competitive with the corresponding bounds 
from FERMI-LAT mentioned above. However, this analysis also has its share of uncertainties. 
The bounds depend on the choice of the propagation parameters of the antiproton and uncertainties 
in the estimates of the astrophysical background. Using a specific choice of propagation parameters 
and a conservative background estimates, it was shown that for $ \mlspone \leq $ 150 GeV, one can rule out 
LSP-chargino mass splittings up to 20 GeV \cite{boehmpamela}.
This result if established conclusively, may nicely complement 
the LHC searches, which are rather insensitive to small LSP-chargino mass splittings. It has also been found that pure wino or wino-like neutralinos are excluded if they are lighter than 450 GeV.

We end this discussion by noting that none of the indirect search experiments are on the same footing as the WMAP experiment. The WMAP experiment probes the standard model of cosmology as a whole. Their  measurement of the DM  relic density is, therefore, part of a much larger canvas (i.e.,  our current understanding of the universe).   

In contrast the above direct and indirect searches - though interesting - have limited goals. More important, the interpretation of their results involve  additional assumptions and astrophysical inputs which cannot be justified rigorously with our present state of knowledge. In our opinion it is best to treat these results as hints rather than conclusive evidences.

\section{Conclusions}

In this paper we have considered several scenarios involving light EW
sparticles (Section 2) which can potentially explain the observed DM relic
density of the universe. Some of these scenarios have been constrained by
the ATLAS and CMS data.

In the most economical model only the LSP is light with mass $\mlspone
\approx m_h/2$, while all other sparticles are heavy. LSP pair
annihilation into the lighter h-scalar is the relic density producing
mechanism. The
corresponding LHC signature in the monojet + $\met$ channel would be rather
hard to detect.

A two sparticle model of the EW dark matter consists of the LSP and a
light R-type slepton (the LRS model). This model has not been tested at
the LHC so far as the R-slepton production cross section is rather tiny. Both the
above models may be easily tested at a suitable $e^+ e^-$ collider.

Another two sparticle model which has already been tested by the search
for slepton pair production is the light L-slepton (LLS) scenario (see
Fig. \ref{slepton}). The above choice $\mlspone \approx m_h/2$ is disfavoured by the
LHC data for $130 < \mslepl < 180$ GeV. Apart from the LSP pair annihilation
into the h-resonance,
LSP - sneutrino (NLSP) co-annihilation could be the main relic density
producing mechanism (see Fig. \ref{slepton}). The co-annihilation strip corresponding
to a small mass difference between the LSP and the L-slepton is far away
from the parameter space sensitive to the current LHC experiments. This
indicates that it will be hard to test this model by the LHC experiments
at least in the near future.

A three particle model consisting of light L and R-type sleptons (LLRS) in
addition to the LSP will have the same LHC signature as the LLS model.
However, due to the R-slepton the DM producing mechanism could be
different depending on the magnitude of mixing in the stau mass matrix
which determines whether the lighter stau or the sneutrino would be the
LSP. The parameter space consistent with the WMAP data  with the choice $
m_{\tilde l_R} =  m_{\tilde l_R}$ is shown in
Fig 4. The regions lying
close to the present exclusion contours may be probed by the next round of
LHC experiments.

We have also considered the LLRS model with larger LSP and slepton masses.
The parameter space consistent with the WMAP data is shown in Fig. \ref{slepton_arg}. The
dominant relic density producing mechanisms are more or less the same as the
ones discussed above.

A model with heavy sleptons but relatively light $\chonepm$ and $\lsptwo$
(LGHS) is another scenario with three light sparticles considered in this
paper. The present LHC constraints are exhibited in Fig \ref{lghs}. The DM
producing mechanism with $\mlspone \approx  m_h/2$ is disfavoured for 
$ 180 < \mchonepm < 290$ GeV. The co-annihilation strip shown in Fig. \ref{lghs} is
also consistent with the WMAP data. Here $\lspone$, $\lsptwo$ and
$\chonepm$ are nearly degenerate. The LHC signature will be very hard to
detect.

The LGRS model with four light sparticles consists of
$\chonepm$, $\lsptwo$ and the R-type slepton. The ATLAS collaboration has
taken
$ m_{\tilde l_R} = (\mlspone + \mlsptwo)/2 $. The resulting LHC and WMAP
constraints are shown in Figs. 1(a), 1(b) and 1(c).
It is expected that the WMAP allowed regions close to the present
exclusion contour may be probed by the
next round of experiments. The CMS collaboration has assumed a more
general relation between the slepton and the gaugino masses (see Section
2). The resulting constraints both from LHC and the observed relic density
can be found in Fig. \ref{lgrscms}.  A large fraction  of the parameter space
consistent with the WMAP data has already been excluded. The regions close
to the exclusion contours may be probed in the near future. Probing the
long coannihilation strips with nearly degenerate R-slepton and the LSP at
the LHC  will be rather challenging.

A similar model comprising of $\chonepm$, $\lsptwo$ and L-type sleptons
(LGLS) have also been considered. The constraints from the ATLAS
collaboration
with the choice $ m_{\tilde l_L} = (\mlspone + \mlsptwo)/2 $ can be found
in Fig. 7(a) of \cite{atlas3l8}. The main relic density producing mechanism
is
the LSP-sneutrino co-annihilation. As a result the mass splitting between
the slepton  and the LSP is also small but not as small as that between
the LSP and the sneutrino. Nevertheless the trilepton signal in this model
would be rather hard to detect.

The model with light electroweak gauginos and both L and R type sleptons
(LGLRS) have not been tested at the LHC. However, we have considered the signatures
of this model in the light gluino scenario (Section 3).

In addition to a light EW sector the gluino could be relatively light
without affecting the predictions for the DM relic density. However, this
may lead to interesting consequences for LHC search. In Table \ref{tab2} we present
the revision of the gluino mass limit for different benchmark scenarios
(see Table \ref{tab1}). These scenarios are consistent with the LHC constraints on
the electroweak sparticles and the observed DM relic density. It may be
recalled that in the mSUGRA model the strongest gluino mass limit for
heavy squarks is about 1 TeV obtained by searches in the jets
+ $\met$ channel.  It is worth stressing that in the LGLS and LGLRS models 
(BP8, BP9 and BP10) the limit may be relaxed by as much as 25\%. The corresponding
limits in the SSD + jets + $\met$ channel, however, are quite competitive
with the strongest  limits currently available.

If the gluinos are just beyond the current reach of the LHC then EW scenarios 
having different DM producing mechanisms, can be distinguished by the 
n-leptons + m-jets + $\met$ signal. This is illustrated by the relative rates of the signal events for
n=0 and n=2 (the SSD signal) measured through the ratio $r_1$ defined in
Table \ref{tab4}. It may be stressed that this ratio involves very little theoretical
uncertainty and can be measured with sufficient accuracy if the
backgrounds are reduced to negligible levels with the accumulation of
data.
On the one hand there are scenarios where the SSD signal is strongly
suppressed and $r_1$ is rather large (see Table~\ref{tab4} second column). On the other extreme
there are the LGLS or the LGLRS scenario where $r_1$ is small (Table \ref{tab4} second column). 
It is worth noting that in all models with either light L or R type sleptons $r_1 < 10$.
If the light slepton is removed the corresponding ratio $r_2$ is much larger (see Table~\ref{tab4} third column).

The LGLS or the LGLRS model can  potentially
remove the alleged discrepancy between the theoretical prediction for
$(g-2)_{\mu}$ in the SM with the data. The parameter space compatible with
LHC, WMAP and $(g-2)_{\mu}$ constraints can be easily located in Fig. 1 of
\cite{endo}. The regions just above the thick black lines indicating $\msnu
\approx \mlspone$ in Figs. 1(a), 1(b) and 1(d) satisfy all the above constraints.

The light slepton scenario as discussed in this paper are intimately connected to novel 
collider
signatures, the observed DM relic density of the universe and enhanced prediction for
the $(g -2)_{\mu}$ of the muon. It is, therefore, extremely important to asses the
prospect of searching the chargino-neutralino and slepton pair production signals at
the LHC 13/14 TeV. A recent analysis has addressed the prospect of  gaugino pair 
production via gauge boson fusion at the LHC 14 TeV experiments \cite{bhaskar}. 

In this context 
it may be noted that in a large parameter space of the EW sector of the MSSM,
not extensively discussed, in this paper,  
the stau is the NLSP. In these regions the electroweak signals dominantly consists 
of tau rich final states \cite{adnabanita}. Using efficient tau tagging at the LHC
one can probe additional parameter spaces via these channels.

{ \bf Acknowledgments : } We thank Prof. Utpal Chattopadhyay (IACS, India) 
and Manimala Chakraborti (IACS, India) for useful discussions.


\end{document}